\documentclass[twocolumn,secnumarabic,amssymb, nobibnotes, aps,graphicx,pre]{revtex4-1}
\usepackage{amssymb}
\usepackage{bm}
\usepackage{graphicx}
\usepackage{amsmath}
\usepackage{color}
\def\1{{\bf 1}}
\def\0{{\bf 0}}
\def \RC {C\!_{{\scriptscriptstyle R}}}
\def \ZM {Z_{{\scriptscriptstyle M}}}
\def \rF { F_{er}^z}
\def \rU {{\cal U}\!_r}

\def \ru {u_r}
\def \rga {\gamma\!_f\,}

\def \D {{\cal D}}
\def \E {{\cal E}}

\def \U {{\cal U}}

\def\nn{\nonumber \\}

\newcommand{\bY}{{\bm Y}}
\newcommand{\bx}{{\bm x}}
\newcommand{\bX}{{\bm X}}
\newcommand{\bu}{{\bm u}}

\newcommand{\bv}{{\bm v}}
\newcommand{\bz}{{\bm z}}
\newcommand{\Ba}{{\bm \alpha}}
\newcommand{\bb}{{\bm \beta}}

\newcommand{\Be}{{\bm \epsilon}}
\newcommand{\bE}{{\bm E}}
\newcommand{\bB}{{\bm B}}
\newcommand{\bA}{{\bm A}}

\newcommand{\Bp}{{\bm p}}

\newcommand{\be}{\begin{equation}}
\newcommand{\ee}{\end{equation}}
\newcommand{\bea}{\begin{eqnarray}}
\newcommand{\eea}{\end{eqnarray}}
\newcommand{\ba}{\begin{array}}
\newcommand{\ea}{\end{array}}
\def\sq{\mbox{\rlap{$\sqcap$}$\sqcup$}}
\newenvironment{proof}[1]{\vspace{5pt}\noindent{\bf Proof #1}\hspace{6pt}}%
{\hfill\sq}
\newcommand{\bp}{\begin{proof}}
\newcommand{\ep}{\end{proof}\par\vspace{10pt}\noindent}
\begin{document}

\title{Simple model of the slingshot effect}

\author{  Gaetano Fiore$^{1,3}$, \  Sergio De Nicola$^{2,3}$ 
   \\    
$^{1}$ Dip. di Matematica e Applicazioni, Universit\`a di Napoli ``Federico II'',\\
Complesso Universitario  M. S. Angelo, Via Cintia, 80126 Napoli, Italy\\         
$^{2}$  SPIN-CNR, Complesso  MSA, Via Cintia, 80126 Napoli, Italy\\
$^{3}$         INFN, Sez. di Napoli, Complesso  MSA,  Via Cintia, 80126 Napoli, Italy}



\begin{abstract}
\noindent
We present a detailed quantitative description of the recently proposed 
``slingshot effect". Namely, we determine a broad 
range of conditions under which the impact  of a very short and 
intense laser pulse normally onto a low-density plasma (or matter  locally completely 
ionized into a plasma by the pulse) causes the expulsion of a bunch 
of surface electrons in the direction opposite to the one of propagation 
of the pulse, and the detailed, ready-for-experiments features of the 
expelled electrons (energy spectrum, collimation, etc). The effect is due to the combined
actions of the ponderomotive force and the huge longitudinal field arising from charge
separation. 
Our predictions are based on estimating 3D corrections to a simple, 
yet powerful plane 2-fluid magnetohydrodynamic (MHD) model 
where the equations to be solved are reduced to
a system of Hamilton equations in one dimension (or a collection of) 
which become autonomous after the pulse has overcome the electrons.
Experimental tests seem to be at hand.
If confirmed by the latter, the effect would provide a new extraction and acceleration mechanism for electrons,
alternative to traditional radio-frequency-based or Laser-Wake-Field  ones.
\end{abstract}

\maketitle

\section{Introduction and set-up}  

Laser-driven Plasma-based Acceleration (LPA) mechanisms were first conceived 
by Tajima and Dawson in 1979 \cite{Tajima-Dawson1979} and have been 
intensively studied since then. In particular, after the rapid development \cite{StriMou85,
PerMou94} of chirped pulse
amplification  laser technology -  making available compact
sources of intense, high-power, ultrashort laser pulses - the Laser Wake Field Acceleration (LWFA)
mechanism  \cite{Tajima-Dawson1979,Gorbunov-Kirsanov1987,Sprangle1988} 
allows to  generate extremely high 
acceleration gradients ($>$1GV/cm) by plasma waves involving
huge charge density variations. Since 2004 experiments  have shown
that LWFA  in the socalled {\it bubble} (or {\it blowout})  regime  can produce electron bunches 
of high quality (i.e. very good collimation and small
energy spread), energies of up to hundreds of MeVs \cite{ManEtAl04,
GedEtAl04,FauEtAl04} 
  or more recently even GeVs  \cite{WanEtAl13,LeeEtAl14}.
This  allows a revolution in acceleration techniques
of charged particles, with a host of potential applications in research
(particle physics, materials science, structural biology, etc.)
 as well as applications in medicine, optycs, etc. 

\begin{figure*}
\begin{center}
\includegraphics[width=8cm]{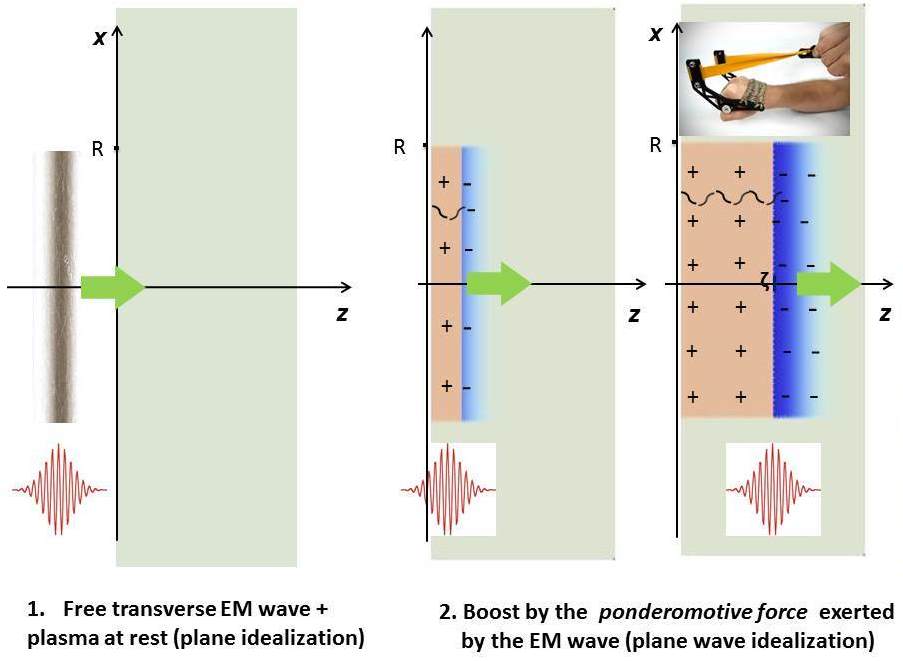} \hfill
\includegraphics[width=8cm]{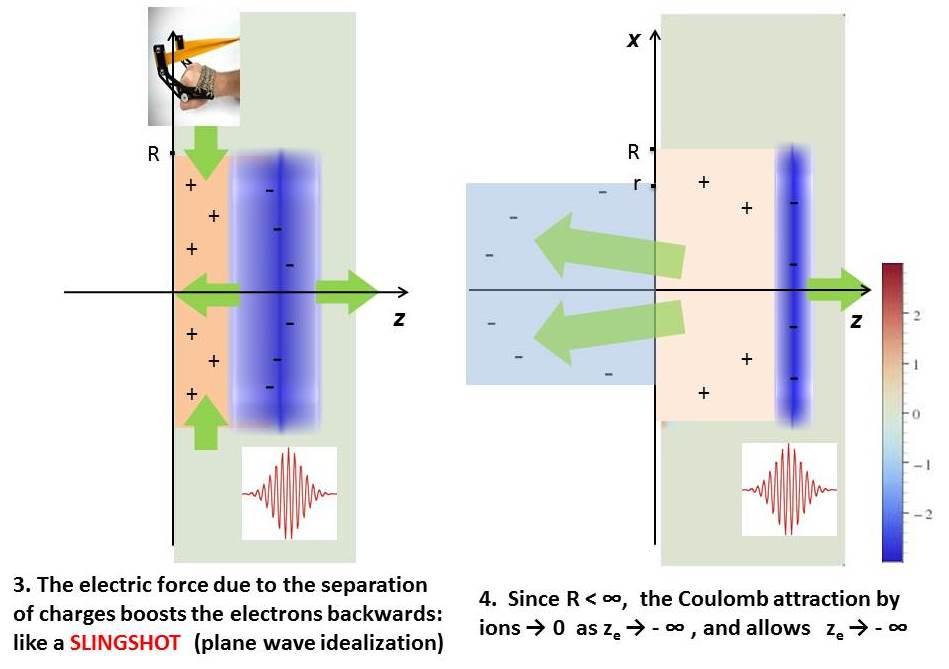}
\end{center}
\caption{Schematic stages of the slingshot effect}
\label{plasma-laser2}      
\end{figure*}
In the LWFA and its variations the laser pulse travelling in the plasma leaves a wakefield of plasma waves behind;
a bunch of electrons (either externally \cite{Irman2007} or self injected \cite{Joshi2006}) can be accelerated ``surfing"
one of these plasma waves and exit the plasma sample just behind the pulse,
in the same direction of propagation of the latter (forward expulsion).
 In Ref. \cite{FioFedDeA14} a new LPA mechanism, named {\it slingshot effect}, has been proposed,
in which a bunch of electrons is expected to be accelerated and expelled {\it backwards} from a 
low-density plasma sample 
shortly after the impact of a suitable ultra-short and ultra-intense laser pulse in the form of a 
pancake normally onto the plasma (see fig. \ref{plasma-laser2}).
The surface electrons (i.e.  plasma electrons in a thin layer just beyond the vacuum-plasma interface) 
first are all displaced forward (with respect to the ions) by the ponderomotive force 
$F\!_p\!:=\!\langle -e(\frac{\bv}c \times \bB)^z\rangle$
generated by the pulse, leaving a layer of ions completely depleted of electrons (here   $\langle\: \rangle$ is 
 the average over a period of the laser carrier wave, $\bE,\bB$ are the electric and magnetic fields,
$\bv$ is the electron velocity, $c$ is the speed of light,  $\hat \bz$ is the direction of propagation
of the laser pulse); $F\!_p$  is positive (negative) while the modulating amplitude $\epsilon_s$ 
of the pulse  respectively grows (decreases). These electrons
are then  pulled back by the longitudinal electric force 
$F_e^z\!=\!-eE^z$ exerted by the ions 
and the other electrons,  and leave the plasma. [In the meanwhile the pulse proceeds deeper into the plasma,
generating a wakefield.]
Tuning the electron density in the range where the
plasma oscillation period $T\!_{{\scriptscriptstyle H}}$\cite{footnote1}
is about twice the pulse duration $\tau$, 
we can make these electrons invert their motion when they are reached by the maximum of $\epsilon_s$, so that the negative part of $F\!_p$  (due to the subsequent decrease of $\epsilon_s$) adds to  $F_e^z$ in  accelerating them backwards; thus the total work 
$W\!=\!\int_0^\tau \! dt \, F\!_p \langle v^z\rangle $ done by the ponderomotive force is  maximal \cite{footnote2}. 
Provided the laser spot size $R$ is sufficiently small a significant part of the expelled electrons will have enough  energy to 
win the attraction by  ions and escape to infinity. 

Very short  $\tau$'s and huge nonlinearities
make   approximation schemes  based on  Fourier analysis and related methods
(slowly varying amplitude approximation, frequency-dependent refractive indices,...)
unconvenient. 
On the contrary,  in the relevant space-time region
a MHD description of the impact  is self-consistent, simple and predictive (collisions are negligible, and 
recourse to  kinetic theory is not needed). 
Here we develop and improve the 2-fluid MHD approach introduced in
 \cite{FioFedDeA14,Fio14JPA} and apply it to determine a broad range of conditions enabling the effect, as well as 
detailed quantitative predictions about it (a brief summary is given in \cite{FioDeN16,Fio16b}). 
In  section \ref{Planewavessetup} we study the  plane 
problem ($R\!=\!\infty$) and show that for sufficiently low density and small times (after the impact)
we can neglect the radiative corrections [back-reaction of the plasma on the electromagnetic (EM) field (\ref{pump})]
and determine the motion of the surface electrons in the bulk by (numerically) solving a single system of 
two coupled first order ordinary differential equations
of Hamiltonian form, if the initial density  $ \widetilde{n_0}$ is step-shaped,  or a collection of such systems,
otherwise;  the role of `time' is played by the light-like coordinate $\xi\!=\!ct-z$. The rough model 
of \cite{FioFedDeA14} considered only step-shaped $ \widetilde{n_0}$ and
was based on neglecting: $F_e^z$ during the forward motion, 
$F_p$ during the backward motion of the electrons; the estimates could be considered reliable only
for very low, unrealistic  $ \widetilde{n_0}$.
Here  $ \widetilde{n_0}$ needs no longer to be  so low, nor  step-shaped, as in \cite{FioFedDeA14},
because we take $F_e^z,F_p$ in due account during the whole motion of the electrons.
In section \ref{3D-effects} we heuristically modify the potential energy 
outside the bulk to account for finite $R$ and
determine a $R$-range  such that 
the motion of the surface electrons coming from  some inner
cylinder  $\rho^2\!\le\! r^2\!<\!R^2$ be (by causality) well approximated by the solution of the correspondingly
modified Hamilton equations;
we then  find which electrons  indeed  escape to infinity and estimate in detail
their final energy spectrum, collimation, total number, charge and energy.  To be specific, 
in section \ref{experiments} we specialize predictions to potential experiments at the FLAME
facility (LNF, Frascati) or the ILIL laboratory (INO-CNR, Pisa).
We welcome 3D simulations and experiments checking these predictions;
the experimental conditions are at hand in many laboratories.
In section \ref{Final} we discuss the results, the conditions for their validity and draw the conclusions.

As a context remark, we recall that relatively simple
2-fluid magnetohydrodynamic models can be used also to describe the complicated
physics of the impact of very intense and short laser pulses on overdense solid targets. If the density gradient
of the target is sufficiently steep, the massive displacement of electrons (induced 
by the ponderomotive force) with respect to ions (named {\it snowplow} in  \cite{SahEtAl13,Sah14}) produces 
a longitudinal electric force which may accelerate also protons or other light ions, either backward
or forward, by the socalled {\it Skin-Layer Ponderomotive Acceleration} \cite{BadEtAl04}  or {\it Relativistically 
Induced Transparency Acceleration} \cite{SahEtAl13,Sah14} mechanisms.

\subsection*{The 2-fluid magnetohydrodynamic framework}

The set-up is as follows.
We assume that the plasma is initially 
neutral, unmagnetized and at rest with electron (and proton) density equal to zero  in the region  \ $z\!<\! 0$. 
We describe the plasma as consisting of a static background fluid of ions (the motion of ions can be neglected
during the short time interval in which the effect occurs)
and a fully relativistic collisionless fluid of electrons, with the ``plasma + EM field" system fulfilling 
the Lorentz-Maxwell and the continuity equations.  We  show 
{\it a posteriori}  that such a MHD treatment 
is self-consistent in the spacetime region of interest.
We denote as \ $\bx_e(t,\bX)$  \ the position at time $t$
of the electrons' fluid element initially located at $\bX\!\equiv\!(X,Y,Z)$, and for each fixed $t$ as
$\bX_e(t,\bx)$ the inverse map [$\bx\!\equiv\!(x,y,z)$]. For brevity,  we refer: to such a
fluid element as to the  ``$\bX$ electrons"; to the fluid elements with arbitrary $X,Y$
and specified $Z$, or  with $\bX$ in a specified
region $\Omega$, respectively as the ``$Z$ electrons" or the ``$\Omega$ electrons''.
We denote as $m,n_e,\bv_e$ the electron mass, Eulerian
density and velocity  and often use the dimensionless fields \ $\bb_e\!\equiv\!\bv_e/c$,
$\bu_e\!\equiv\!\Bp_e/mc\!=\!\bb_e/\sqrt{1\!-\!\bb_e^2}$, 
$\gamma_e\!\equiv\!1/\sqrt{1\!-\!\bb_e^2}\!=\!\sqrt{1\!+\!\bu_e^2}$. \
The equations of  motion are 
\be
\ba{l}
\displaystyle\frac{d \Bp_e}{dt}\!=\!-e\left(\bE \!+\!\frac{\bv_e}c \wedge \bB \right), \\[10pt]
\partial_t \bx_e(t,\bX)=\bv_e\!\left[t,\bx_e(t,\bX) \right] 
\ea  \label{lageul}
\ee
in CGS units ($d/dt\!\equiv\!\partial_t\!+\!\bv_e\!\cdot\!\nabla_x$ is the electrons' material derivative)
and  the initial conditions are $\Bp_e(0,\bX)\!=\!\0$,  
$\bx_e(0,\bX)\!=\!\bX $ for $Z\!\ge\! 0$. 
The Lagrangian fields depend on  $t,\bX$, rather than on $t,\bx$, and are distinguished
 by a tilde, e.g. $\widetilde{n_e}(t,\!\bX)=n_e[t,\!\bx_e(t,\!\bX)]$.
The continuity equation $d n_e/dt\!+\!n_e\nabla_x\!\cdot \bv_e=0$ \
follows from the local conservation of the number of electrons, which amounts to
\be
\tilde n_e(t,\bX)\, \det\!\left(\frac {\partial \bx_e} {\partial \bX}\right)
= \widetilde{n_0}(\bX)\equiv \tilde n_e(0,\bX). \label{n_eg}
\ee
We assume that $ \widetilde{n_0}$ is independent of $X,Y$ and, as said, vanishes 
if $Z\!<\!0$;  also as a warm-up to more general $Z$-dependence, we start by studying the case that it is
constant in the region $Z\!\ge\! 0$: $ \widetilde{n_0}(Z)\!=\!n_0\theta(Z)$, 
where $\theta$ is the Heaviside step function.
 We consider a purely transverse EM pulse in the form of a pancake with cylindrical symmetry around the $z$-axis,
propagating in the positive ${\hat{\bm z}}$ direction and hitting the plasma surface \ $z\!=\!0$ at $t\!=\!0$.
We schematize the pulse as a free plane pulse multiplied by a ``cutoff'' function
$\chi\!_{{\scriptscriptstyle R}}(\rho)$ which is approximately equal to 1 for $\rho\!\equiv\sqrt{\!x^2\!+\!y^2}\!\le\! R$ 
and rapidly goes to zero for $\rho\!>\! R$ (with some  finite  radius $R$,  see fig.
\ref{plasma-laser2}-1)
\be
\bE^{{\scriptscriptstyle\perp}}\!(t,\!\bx)=\Be\!^{{\scriptscriptstyle\perp}}\!(ct\!-\!z)\,
\chi\!_{{\scriptscriptstyle R}}(\rho),\qquad  \bB^{{\scriptscriptstyle\perp}}=
{\hat{\bm z}}\!\times\!\bE\!^{{\scriptscriptstyle\perp}}                                                  \label{pump}
\ee
[in particular we consider 
$\chi\!_{{\scriptscriptstyle R}}(\rho)\!\equiv\!\theta(R\!-\!\rho)$];
the  `pump'  $\Be^{{\scriptscriptstyle\perp}}\!(\xi)$  vanishes
outside some finite interval  $0\!<\!\xi\!<\!l$ \cite{footnote2bis}.

\section{Plane wave idealization}
\label{Planewavessetup}

\begin{figure}
\begin{center}
\includegraphics[width=8cm]{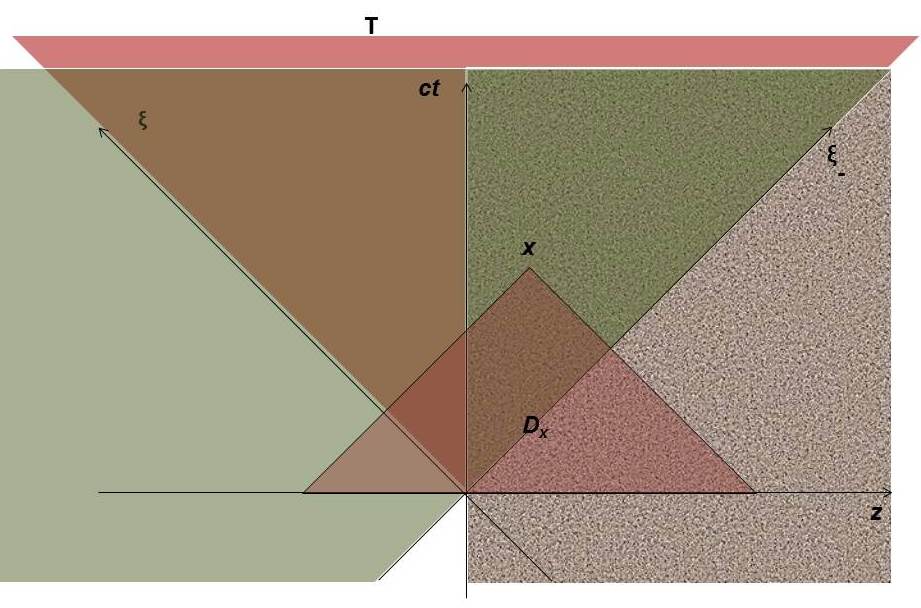}
\end{center}
\caption{Past  (light brown) and future (purple) causal cones $D\!_{x},T$;
supports of $\bA^{{\scriptscriptstyle\perp}}$ (light green) and $ \widetilde{n_0}(z)$
(anthracite).}
\label{DT}       
\end{figure}

In the plane problem  ($R\!=\!\infty$) 
the invertibility of  $\bx_e\!:\!\bX\!\mapsto\! \bx$ for all fixed $t$ amounts
to $z_e( t\!,Z)$ being strictly increasing  with respect to
$ Z$ for   all $t$.
Eq. (\ref{n_eg}) becomes
\be
\ba{l}\left[\tilde n_e \frac{\partial z_e}{\partial Z}\!\right]\!\!( t,\!Z) \!=\! \widetilde{n_0}( Z )
\quad\Leftrightarrow \quad n_e(t,\!z)\!=\! \widetilde{n_0}\!\left[Z_e\!(\!t,\!z\!)\right]\frac{\partial Z_e}{\partial z}(t,\!z).    
\ea\label{n_e}
\ee
Regarding ions as immobile, the Maxwell equations imply  \cite{Fio14JPA} that the longitudinal component of the electric field
is related to  $\widetilde{N}(Z)\!\equiv\!\int^Z_0\!dZ'\, \widetilde{n_0}(Z')$ (the number  of electrons per unit surface
in the layer $0\!\le\!Z'\!\le\! Z$) by
\be
E^z(t, z)\!=\!4\pi e \big\{
\widetilde{N}(z)\!-\! \widetilde{N}[Z_e (t, z)] \big\}.
\quad   \label{elFL}{}
\ee
We partially fix the gauge \cite{Fio14JPA} imposing that the transverse  
(with respect to ${\hat{\bm z}}$) vector potential itself is independent of $x,y$, and hence is the physical observable 
$\bA\!^{{\scriptscriptstyle\perp}}\!(t,z)\!=\!-\!\int^{t}_{-\!\infty }\!\!dt' c\bE^{{\scriptscriptstyle\perp}}\!(t'\!,z)$;
then 
 \ $c\bE^{{\scriptscriptstyle\perp}}\!=\!-\partial_t\bA\!^{{\scriptscriptstyle\perp}}$,
\ $\bB\!=\!\bB^{{\scriptscriptstyle\perp}}\!=\!{\hat{\bm z}}
\!\wedge\!\partial_z\bA\!^{{\scriptscriptstyle\perp}}$.
As known, the  transverse component  of the Lorentz equation (\ref{lageul})$_1$
implies \
$\Bp_e^{{\scriptscriptstyle \perp}} \!-\!\frac ec\bA^{{\scriptscriptstyle \perp}}\! \!\!=$const
on the trajectory of each electron; \ this is zero  at $t\!=\!0$, \  hence \
$\Bp_e^{{\scriptscriptstyle \perp}}\!=\!mc\bu_e^{{\scriptscriptstyle \perp}} \!=\!e\bA\!^{{\scriptscriptstyle \perp}}\!/c$.
Hence $\bu{{}}_e^{{\scriptscriptstyle\perp}}$ is determined in terms of $\bA\!^{{\scriptscriptstyle\perp}}$.
As in \cite{Fio14JPA}, we introduce the positive-definite field
\be
s_e\!\equiv\!\gamma_e\!-\! u^z_e,                                                       \label{defs_e}
\ee
which we name electron {\it s-factor}.
\ $u_e^z,\gamma_e,\bb_e^{{\scriptscriptstyle\perp}},\beta_e^z$ are recovered from $\bu_e^{{\scriptscriptstyle\perp}},s_e$ through the formulae (44) of \cite{Fio14JPA}:
\bea
\gamma_e\!=\!\frac {1\!+\!\bu_e^{{\scriptscriptstyle\perp}}{}^2\!\!+\!s_e^2}{2s_e}, \quad\: u_e^z\!=\!\frac {1\!+\!\bu_e^{{\scriptscriptstyle\perp}}{}^2\!\!-\!s_e^2}{2s_e}, \quad\:  \displaystyle\bb_e\!=\! \frac{\bu_e}{\gamma_e}. \quad \label{u_es_e}
\eea
Remarkably, all of (\ref{u_es_e}) are {\it rational functions} of $\bu_e^{{\scriptscriptstyle\perp}},s_e$ (no square roots appear). 
Moreover,  fast oscillations of $\bu_e^{{\scriptscriptstyle\perp}}$ affect $\gamma_e,u_e^z$ but not
$s_e$ [see the comments after (\ref{heq2})]. For these reasons
it is convenient to use $\bu_e^{{\scriptscriptstyle\perp}},s_e$ instead of $\bu_e^{{\scriptscriptstyle\perp}},u_e^z$ as independent unknowns.
The evolution equation  of $s_e$ (difference of the ones of $\gamma_e,u_e^z$; the former is the scalar product of (\ref{lageul})$_1$  with $\Bp_e/\gamma_e m^2c^2$) reads
 \be
\gamma_e\frac{d  s_e}{dt} \!=\!\frac{e E^{{\scriptscriptstyle z}}}{mc} s_e\!+\! (\partial_t\!+\!c\partial_z)\bu_e^{{\scriptscriptstyle\perp}}{}^2.                        \label{bla'}
\ee
The Maxwell equation for $\bA\!^{{\scriptscriptstyle\perp}}$ takes the form \
$(\partial_0^2\!-\!\partial_z^2)\bA\!^{{\scriptscriptstyle\perp}}
\!+\! \bA\!^{{\scriptscriptstyle\perp}}4\pi e^2n_e/mc^2\gamma_e\!=\!0$; \ 
eq. (\ref{pump}) with $R\!=\!\infty$ implies $\bA\!^{{\scriptscriptstyle\perp}}(t\!,z)\!=\!
\Ba\!^{{\scriptscriptstyle\perp}}(ct\!-\!z)$ for $t\!\le\!0$, where 
$\Ba^{{\scriptscriptstyle \perp}}\!(\xi)\!\equiv\!-\!\!\int^{\xi}_{-\infty }\!\!d\xi' \Be^{{\scriptscriptstyle\perp}}\!(\xi')$.
Using the Green function of the D'Alembertian $\partial_0^2\!-\!\partial_z^2$, abbreviating $x\!\equiv\!(t,z)$, these equations can
be equivalently reformulated as the integral equation (42) of \cite{Fio14JPA}
\bea
\bA\!^{{\scriptscriptstyle\perp}}(t\!,z)-
\Ba\!^{{\scriptscriptstyle\perp}}(ct\!-\!z)=-\!\!
\displaystyle\int_{D\!_{x}\cap T}\!\!\!\!\!\! dt' dz'\left[\frac{2\pi e^2n_e}{mc\gamma_e} \bA\!^{{\scriptscriptstyle\perp}}\right]\!\!(x')  
 \qquad \label{inteq1}  \\[8pt]
D\!_{x}\!\equiv\!\{ 
(t',z')\:|\: {t}' \!\le\! t,
\, |z\!-\!z'|\!\le\!{ct}\!-\!{ct}' \}, \quad  T\!\equiv\!\{ x\:|\:  |z|  \!<\! ct \}
\nonumber
\eea
The past, future causal cones $D\!_{x},T$, the supports of $\bA^{{\scriptscriptstyle\perp}}, \widetilde{n_0}(z)$,
 and their intersections are shown in fig. \ref{DT}.
For  $t\!<\!0$ \ $D\!_{x}\!\cap\! T$ \ is empty, and the right-hand side of (\ref{inteq1})$_1$
is zero, as it must be. Below we shall analyze the consequences of neglecting it also 
for small $t$, and determine the range of validity of such an approximation. 

\subsection{Motion of the electrons}
\label{planemotion}

\bea
\ba{l}\mbox{Let}\qquad \displaystyle\hat\bu^{{\scriptscriptstyle \perp}}(\xi)\!\equiv\! e\Ba^{{\scriptscriptstyle \perp}}(\xi)/mc^2,
\qquad v(\xi )\!\equiv\!\hat\bu^{{\scriptscriptstyle \perp}2}(\xi),\qquad\\[6pt]
\qquad\quad\:\: F_e^z(z,\!Z)\!\equiv\!-4\pi e^2\left\{
\widetilde{N}(z) \!-\! \widetilde{N}(Z)\right\}. 
 \ea \label{definitions}
\eea
$ \widetilde{F_e^z}(t,\!Z)\!\equiv\!F_e^z[z_e\!(t,\! Z),\!Z]$ is the longitudinal electric force acting on the $Z$ electrons
at time $t$; 
it is conservative, as it depends on $t$ only through $z_e(t,\!Z)$.
The approximation $\bA\!^{{\scriptscriptstyle\perp}}(t\!,z)\!=\!\Ba\!^{{\scriptscriptstyle\perp}}(ct\!-\!z)$  implies 
$\bu_e^{{\scriptscriptstyle \perp}}(t\!,z)\!=\!\hat\bu^{{\scriptscriptstyle \perp}}(ct\!-\!z) $, and  the last term of (\ref{bla'}) vanishes.
Replacing (\ref{elFL}) in the Lagrangian version of  (\ref{bla'}), \  
we find for each $Z\!\ge\! 0$ the equation \  $\tilde\gamma_e\partial_t \tilde s_e=-\tilde s_e \widetilde{F_e^z}/mc$. The initial condition is
 $\tilde s_{e}(0,Z)\!\equiv\! 1$. The other equation  to be solved is (\ref{lageul})$_2$  
\ with the initial condition \
$\bx_e(0,\bX)\!=\!\bX$. By (\ref{u_es_e}) one is thus led to the Cauchy problems (parametrized by $Z\!\ge\! 0$)
\bea
&&
\partial_t \frac{z_e\!-\!Z}c=
\frac{1\!+\!v[ct\!-\!z_e(t,\!Z)]\!
\!-\!\tilde s_e^2} {1\!+\!v[ct\!-\!z_e(t,\!Z)]\!+\!\tilde s_e^2 },
\quad
\partial_t \tilde s_e=\displaystyle\frac {-\tilde s_e\!}{\tilde \gamma_e mc} 
\widetilde{F_e^z},\qquad\label{eq1} \\[10pt]
&&  \: z_e(0,Z)\!-\!Z\!=\!0,  \qquad\qquad
 \tilde s_{e}(0,Z)\!=\! 1.  \qquad\qquad\qquad\qquad  \label{eq2}
\eea
$ \bx_e^{{\scriptscriptstyle\perp}}(t,\!\bX)$ is 
obtained  from the solutions of (\ref{eq1}-\ref{eq2}) using (\ref{lageul}), (\ref{u_es_e}):
\be
\bx^{{\scriptscriptstyle \perp}}_e(t,\!\bX )=\bX^{{\scriptscriptstyle \perp}}
\!+\!\! \displaystyle\int^{t}_{0}\!\!
dt{}'\, c\bb_e^{{\scriptscriptstyle \perp}}[t{}', z_e(t{}'\!,\!Z)].\qquad \label{inteq0}
\ee
For all fixed $Z$ the map $t\mapsto \tilde\xi(t,\! Z )\!\equiv\! ct\!-\!z_e(t\!,\!Z)$
is invertible, because the speed of electrons is always smaller than $c$. 
We can simplify (\ref{eq1}) by the change of variables 
 $(t,\! Z )\mapsto \big(\xi,Z\big)$, making the argument of $v$ an independent variable.
Denoting the dependence on $\big(\xi,Z\big)$  by a caret 
[e.g. $\hat s(\xi,\! Z )\!=\!\tilde s_e(t,\! Z )$] and introducing the displacement from the
initial position \ $\hat\Delta(\xi,\! Z)\!\equiv\!\hat z_e(\xi,\!Z)\!-\!Z$, \ we find  \
$\partial_\xi=(\tilde\gamma_e/c\,\tilde s_e)\partial_t$, and  (\ref{eq1}) becomes
\bea
&&
\hat \Delta'=\displaystyle\frac {1\!+\!v}{2\hat s^2}\!-\!\frac 12,
\qquad 
\hat s'=\frac{4\pi e^2}{mc^2}\left\{\!
\widetilde{N}[\hat\Delta\!+\!Z] \!-\! \widetilde{N}(Z)\!\right\}\quad \label{heq1} 
\eea
(the prime means differentiation  with respect to $\xi$). For 
$\xi\!\le\!0$ \ $v(\xi)\!\equiv\!0$, $\hat\Delta,\hat s$ remain constant, 
and we can replace the  initial conditions 
 $\hat\Delta(-Z,\!Z)\!=\!0$, $\hat s(-Z,\!Z)\!=\! 1$ by 
\bea
&&  \: \hat \Delta(0,\!Z)\!=\!0,  \qquad\qquad
 \hat s(0,\!Z)\!=\! 1. \qquad\qquad\qquad\qquad  \label{heq2}
\eea
An alternative derivation of  (\ref{heq1}-\ref{heq2}) with a deeper
insight on the role of the $s$-factor is given in \cite{Fio16c}.
In the zero density limit\ $\widetilde{N}(Z)\!\equiv\!0$, 
$ \hat s\!\equiv\!1$,  (\ref{heq1}-\ref{heq2}) is integrable, and all unknowns
are determined explicitly from $\Be^{{\scriptscriptstyle\perp}}$ 
 \cite{Fio14JPA,Fio14}). As $v\!\ge\! 0$, even if $\Be^{{\scriptscriptstyle\perp}}\!,\bu^{{\scriptscriptstyle \perp}}\!,v$ 
oscillate fast with $\xi$, 
integrating (\ref{heq1}) makes relative oscillations of $\hat\Delta$ much smaller  than those of 
$v$ and  those of $\hat s$ much smaller than the former; hence, $\hat s$ is practically
smooth, see e.g. fig. \ref{graphs}. 
Setting \ $q\!\equiv\!\hat\Delta$,   $p\!\equiv\!-\hat s$, \   for each fixed $Z$
(\ref{heq1}) are the Hamilton equations (with `time' $\xi$) $q'=\partial \check H/\partial p$, $p'=-\partial \check H/\partial q$ of a system with Hamiltonian $\check H(q,p,\xi;Z)\!\equiv\!H(q,-p,\xi;Z)$, 
\bea
\ba{l}
H\!(\Delta,\!s,\!\xi;\!Z)\equiv \gamma(s,\xi)+ \U(\Delta;\!Z),\quad\gamma(s,\xi)\!\equiv\! \displaystyle\frac{s^2\!\!+\!1\!+\!v(\xi)}{2s},\\[8pt]
 \U(\Delta;\!Z)\!\equiv\!\frac{4\pi e^2}{mc^2}\left[
\widetilde{{\cal N}}\!(Z \!+\!\Delta) \!-\!\widetilde{{\cal N}}\!(Z)\!-\! \widetilde{N}\!(Z)\Delta\right],\\[8pt]
 \widetilde{{\cal N}}(Z)\equiv
\int^Z_0\!\!\!dZ'\,\widetilde{N}(Z')\!=\!\int^{Z}_0\!\!\!dZ'\, \widetilde{n_0}(Z')\, (Z\!-\!Z').
\ea                                \label{hamiltonian}
\eea
Defining $\U$  we have fixed the free additive constant so that  $\U(0,\!Z)\!\equiv\! 0$ for each $Z$;
\ $H\!-\!\sqrt{1\!+\!v}$ \ is positive definite. Below we shall abbreviate \ $P(\xi;Z)\!\equiv\!\big(\hat \Delta(\xi;Z),\hat s(\xi;Z)\big)$.

The right-hand side of (\ref{heq1})$_2$ is an increasing function of $\hat\Delta$, because so is $\widetilde{N}(Z)$.
As  $v(\xi )$ \ is zero for $\xi\!\le\!0$ and positive  for small $\xi\!>\!  0$,  then so are also
$\hat\Delta(\xi,\!Z)$ and $\hat s(\xi,\!Z)\!-\!1$. \  Both keep increasing 
until $\hat\Delta$ reaches a positive maximum $\hat\Delta(\bar\xi,Z)$ 
at the $\xi\!=\!\bar\xi(Z)\!>\!0$ such that 
\be
\hat\Delta'(\bar\xi,Z)=0 \quad\Leftrightarrow\quad \hat s^2(\bar\xi,\!Z)\!=\!1\!+\!v(\bar\xi)
\label{defbarxi}
\ee
   (note that $\bar\xi\!<\!l$ if $v(l)\!=\!0$). We shall denote as $\zeta\!\equiv\!\hat\Delta[\bar\xi(0),0]$
the maximum penetration of the $Z\!=\!0$ electrons.
For $\xi\!>\!\bar\xi(Z)$  $\hat\Delta$ starts decreasing; $\hat s$ reaches a maximum at the $\xi\!=\!\xi_r(Z)$ such that $\hat\Delta(\xi_r,\!Z)\!=\!0$ (i.e. at $\xi\!=\!\xi_r(Z)$ the $Z$ electrons have regained their initial $z$). Both decrease for $\xi\!>\!\xi_r(Z)$, until  $\hat s$ becomes so small, and the right-hand side of  (\ref{heq1})$_1$ so large, that first
$\hat\Delta$, and then $\hat s\!-\!1$, are forced to abruptly grow again to positive values. This prevents $\hat s$ to vanish anywhere, consistently with (\ref{defs_e}). 
In $\xi$-intervals where $v(\xi)\!\equiv\!v_c\!\equiv$const,  $H$ is conserved, and all trajectories $P(\xi;Z)$ in phase space (paths)
are level curves $H(\Delta,s;Z)\!=\!h(Z)$, 
above the line $s\!=\!0$, integrable by quadrature \cite{Fio16}.  
For $Z\!=\!0$ the paths are unbounded with $\hat\Delta(\xi,0)\!\to\!-\infty$  as $\xi\!\to\!\infty$. For $Z\!>\!0$ the paths are cycles around the only critical point $C\!\equiv\!(\Delta,s)\!=\!(0,\sqrt{1\!+\!v_c})$ (a center); therefore for $\xi\!\ge\! l$ $v(\xi)\!=\!v(l)$, and these solutions are periodic. There exists a $Z_b\!>\!0$ such that:
the paths $P(\xi;Z)$ with $Z\!<\!Z_b$  cross the $\hat\Delta\!=\!-Z$ line twice, i.e. 
go out of the bulk and then come back into it; the path  $P(\xi;Z_b)$ is tangent to this line in the point
 $(\hat\Delta,\hat s)\!=\!(-Z_b,\sqrt{1\!+\!v(l)})$ (where $\hat\Delta'\!=\!0$); the paths $P(\xi;Z)$ with  $Z\!>\!Z_b$ 
do not cross this line. 
For $Z\!\le\!Z_b$ let $\xi_{ex}(Z)$ be the first positive solution of the equation \ $\hat\Delta(\xi,\!Z)\!=\!-Z$,
i.e. at $\xi\!=\!\xi_{ex}(Z)$ the $Z$ electrons exit the bulk:
\bea
\hat z_e\left[\xi_{ex}(Z),\!Z\right]=0.
\label{defxiex}
\eea
The function $\xi_{ex}(Z)$ is strictly increasing if 
$\partial_{{\scriptscriptstyle Z}}\hat z_e\!>\!0$. 

For any family $P(\xi;Z)$  of solutions of  (\ref{heq1}-\ref{heq2}) 
let 
\bea
&&\hat u^z\!\equiv\!\displaystyle\frac {1\!+\!v\!-\!\hat s^2}{2\hat s},\qquad \qquad
\hat \gamma\!\equiv\!\displaystyle\frac {1\!+\!v\!+\!\hat s^2}{2\hat s} ,\nn
&& \hat\bx_e(\xi,\!\bX)=\bX+  \hat \bY\!(\xi,\!Z),\qquad
 \hat \bY\!(\xi,\!Z)\!\equiv\!\!\displaystyle\int\limits^\xi_0\!\!\! dy \,\frac{\hat\bu(y,\!Z)}{\hat s(y,\!Z)},
\qquad   \label{defYXi}      \\
&& c\hat t(\xi,Z)\!=\!Z\!+\!\hat \Xi(\xi,\!Z),\quad  \hat \Xi(\xi,\!Z)\!\equiv\!\!\displaystyle\int^\xi_0\!\!\!\!\!  dy\, \frac{\hat \gamma(y,\!Z)}{\hat s(y,\!Z)}\!=\!
\xi  \!+\! \hat \Delta(\xi,\!Z)    \nonumber                        
\eea
(note that $\hat Y^z\!=\!\hat \Delta$). The so defined $\hat\bu,\hat\gamma,\hat \bx_e$ 
are the solutions - expressed as functions of $\xi,\bX$ - of all equations and initial 
conditions \cite{footnote3}.
Note that $\hat\bx_e,\hat t$ can be obtained also solving  the  system of functional equations  
\be
\xi=ct\!-\!z,  \qquad  
\hat \Xi(ct\!-\!z,\!Z)\!=\!ct\!-\!Z,  \qquad  \bx-\bX=\hat \bY(ct\!-\!z,\!Z)                         \label{functeq}
\ee
[by (\ref{defYXi}) the second is actually equivalent to the $z$-component of the third] with respect to $t,\bx$.  Clearly \ 
$\hat \Xi(\xi,\!Z)$ \ is strictly increasing and invertible with respect to $\xi$ for all fixed $Z$. 
Solving (\ref{functeq}) with respect to  $\xi,\bx$ (resp. $\xi,\bX$)  as functions of  $t,\bX$
(resp. of $t,\bx$) and replacing the results in $\hat\bu,\hat\gamma,\hat s,...$ 
one obtains  the solutions in the Lagrangian   (resp. Eulerian) description:
in particular one finds (generalizing \cite{Fio14JPA})
\bea
\ba{l}
 \tilde\xi(t,Z)\!=\!
\hat \Xi^{{{\scriptscriptstyle -1}}}\!(ct\!-\!Z,\!Z),\quad
\bx_e(t,\!\bX\! )\!=\!\bX\!\!+\!\hat \bY\!\!\left[\tilde\xi(t,Z),\!Z\!\right]\!,\\
z_e(t\!,\!Z)\!=\!Z+\hat \Delta\!\left[\tilde\xi(t,Z),Z\!\right]\!\!=\! ct\!-\!\tilde\xi(t,Z), 
           \\
 \tilde s_e(t,\!Z) \!\equiv\!  \hat s\!\left[\tilde\xi(t,Z),Z\right], \quad
\tilde\bu_e(t,\!Z )\!=\!\hat\bu\!\!\left[\tilde\xi(t,Z),\!Z\!\right]\!,\\[8pt]
\bX_e^{{\scriptscriptstyle \perp}}(t,\bx)=\bx^{{\scriptscriptstyle \perp}} \!-\!
\hat \bY^{{\scriptscriptstyle \perp}}\!\left[ct\!-\!z,\!Z_e(t,\!z)\right], \\[8pt]
\bu_e(t,\!z )\!=\!\hat\bu\!\left[ct\!-\!z,\!Z_e(t,\!z)\right]\!.   
\ea \qquad  \label{sol}   
\eea
Indeed, it is straightforward to check that 
 $\big(z_e(t,Z),\tilde s_e(t,Z)\big)$  is the solution of (\ref{eq1}-\ref{eq2}) and 
$\Bp_e(t,\bx)\!\equiv\!mc \bu_e(t,\!z )$, \ $\bx_e(t,\!\bX )$ \ of the PDE's (\ref{lageul}) with the initial conditions 
$\Bp_e(0,\bX)\!=\!\0$, $\bx_e(0,\bX)\!=\!\bX $ for $Z\!\ge\! 0$.

From (\ref{defbarxi}),  (\ref{defxiex}), (\ref{defYXi})$_5$,   the times of maximal penetration and 
of expulsion of the $Z$ electrons are 
\be
\bar t(Z)\!=\!\frac{Z \!+\! \hat \Xi(\bar\xi,\!Z)}c, \qquad t_{ex}(Z)\!=\!\frac{Z \!+\! \hat \Xi(\xi_{ex},\!Z)}c.
\label{defbart_tex}
\ee
Deriving (\ref{sol})  and  the
identity \ $y\!\equiv\!\hat\Xi\!\left[\hat\Xi^{-1}(y,\!Z),\!Z\right]$ we obtain a few useful relations, e.g.
\bea
\frac{\partial \hat\Xi^{{{\scriptscriptstyle -1}}}}{\partial Z}
\!=\! \left.\frac{-\hat s}{\hat\gamma}\frac{\partial \hat\Delta}{\partial Z}\right|_{\xi=\hat\Xi^{-1}\!(y,Z)}\!\!\!\!\!\!,
\quad 
\frac{\partial z_e}{\partial Z}
\!=\! \left.\frac{\hat s}{\hat\gamma}\frac{\partial \hat z_e}{\partial Z}\right\vert_{\xi=\hat \Xi^{-1}\!(ct\!-\!Z,Z)}\!\!\!\!\!\! ,\qquad \nn
 \frac{\partial Z_e}{\partial z}
=\left.\frac{\hat\gamma}{\hat s\partial_{{\scriptscriptstyle Z}}\hat z_e}\right\vert_{(\xi,Z)=\big(ct\!-z,Z_e(t,z)\big)}\!\!\!.
\qquad\qquad\qquad\qquad\label{invZtoz}
\eea
By (\ref{invZtoz}), $\partial_{{\scriptscriptstyle Z}}\hat z_e\!\equiv\!1\!+\!\partial_{{\scriptscriptstyle Z}}\hat\Delta\!>\!0$ is thus a necessary  and sufficient condition for
the invertibility of the maps \ $z_e\!:\!Z\!\mapsto\! z$,  $\bx_e\!:\!\bX\!\mapsto\!  \bx$ \ (at fixed $t$), justifying the hydrodynamic description of the plasma adopted so far and the presence of the inverse function $Z_e(t,\!z)$  in (\ref{sol}). 
Finally, from  (\ref{n_e}),  (\ref{invZtoz}) we find also
\bea
 &&  n_e(t,z)\!=\! 
\widetilde{n_0}\!\left[Z_e(t,\!z)\right]\left.\frac{\hat\gamma}{\hat s\,\partial_Z\hat z_e}\right\vert_{(\xi,Z)=
\big(ct\!-z,Z_e(t,z)\big)}\!\!.     \qquad    \label{expln_e}
\eea

We can test the range of validity of the approximation $\bA\!^{{\scriptscriptstyle\perp}}(t\!,z)
\!=\!\Ba\!^{{\scriptscriptstyle\perp}}(ct\!-\!z)$ by showing that the latter makes the modulus of the right-hand side of  (\ref{inteq1}) 
 much smaller than $\alpha\!^{{\scriptscriptstyle\perp}}(ct\!-\!z)$  on \
$D\!\equiv\!\{ (t,z)\:|\:  0\!\le\!ct\!-\!z\!\le\!\xi_{ex}(\ZM), \, 0\!\le\!ct\!+\!z\!\le\! \xi_{ex}(\ZM)\}$ ($\ZM$ is defined below), or equivalently [multiplying by $e/mc^2$ and
using (\ref{expln_e})]
\bea
\mbox{for $x\!\equiv(t,z)\!\in D$} \qquad \left|\delta \bu\!^{{\scriptscriptstyle\perp}}\!(t,z)
 \right|\ll |\bu\!^{{\scriptscriptstyle\perp}}\!(ct\!-\!z)|,  \qquad   \label{condapprox}\\
\delta \bu\!^{{\scriptscriptstyle\perp}}\!(t,z)\! \equiv\!\! \displaystyle\int_{D\!_{x}\cap T}\!\!\!\!\!\! \!\!\!\!\!\! 
dt' dz'\frac{2\pi e^2 \widetilde{n_0}[Z_e(t',\!z')]  \bu\!^{{\scriptscriptstyle\perp}}\!(ct'\!-\!z')  }
{mc\, [\hat s\, \partial_{{\scriptscriptstyle Z}} \hat z_e ]_{(\xi,Z)=\big(ct'\!-z',Z_e(t',z')\big)}  };\qquad \nonumber
\eea
actually, it suffices to check this inequality on the worldlines of the expelled electrons.

\subsection{Auxiliary problem: constant initial density}
\label{Auxiliary}

\begin{figure}
\begin{center}
\includegraphics[width=7cm]{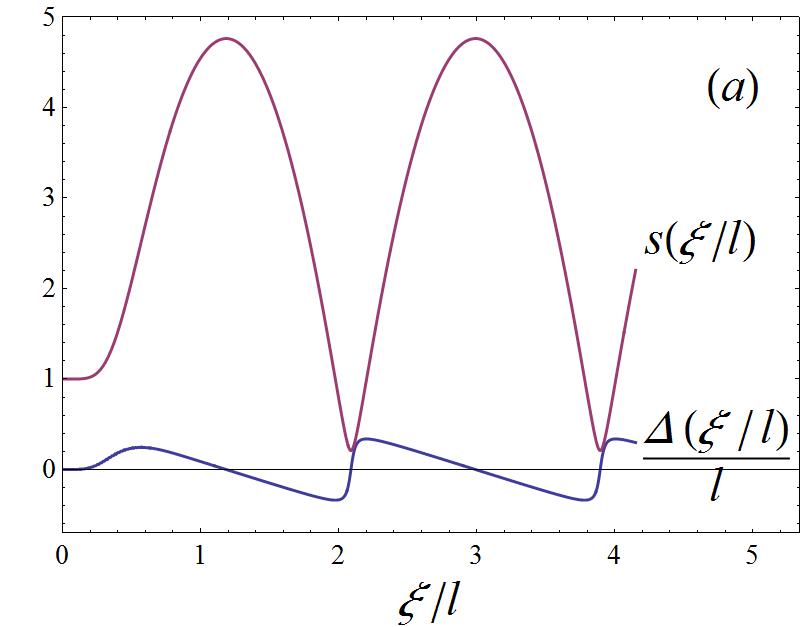} 
\includegraphics[width=6cm]{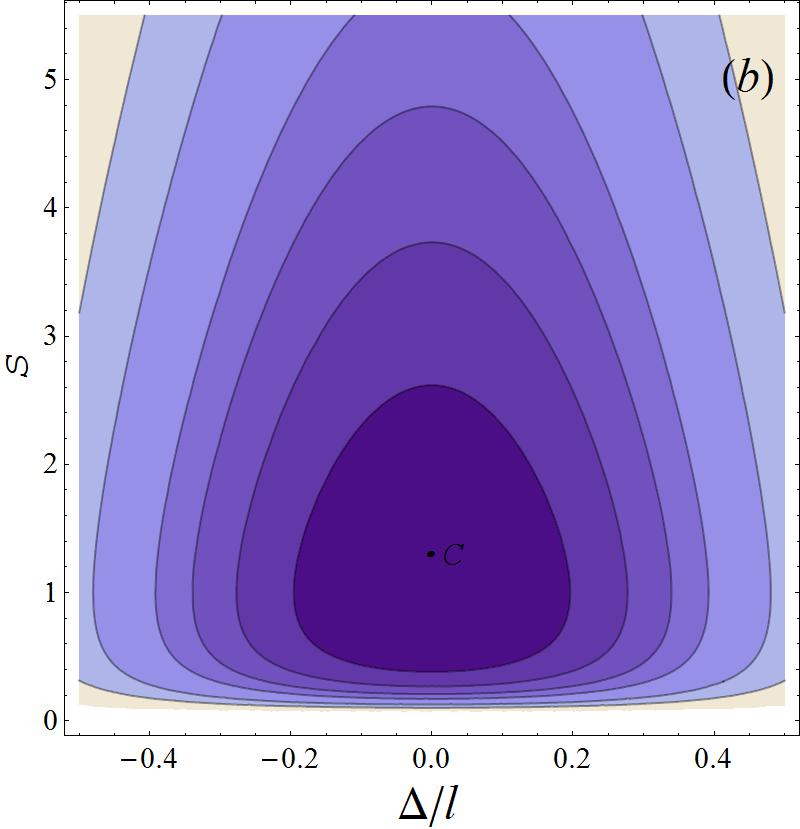}
\end{center}
\caption{(a) Solution of (\ref{e1}) for $Ml^2\!=\!26$ and the $v(\xi)$ as in section
\ref{experiments} of average intensity $I\!=\!10^{19}$W/cm$^2$.
(b) Paths $P(\xi;Z)$  around the center $C$ for $Ml^2\!=\!26$, $v_c\!=\!0$.}
\label{pqCyclesM26}      
\end{figure}

\begin{figure}
\includegraphics[width=7.88cm]{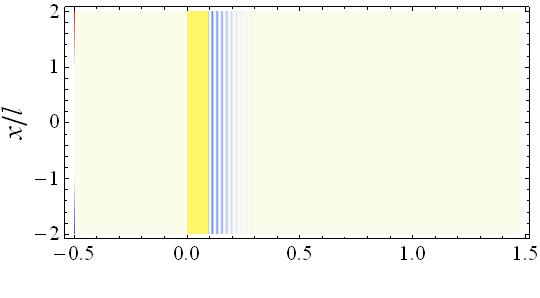} \includegraphics[width=.62cm]{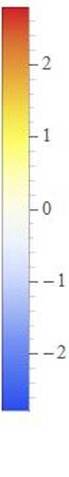}
\includegraphics[width=7.88cm]{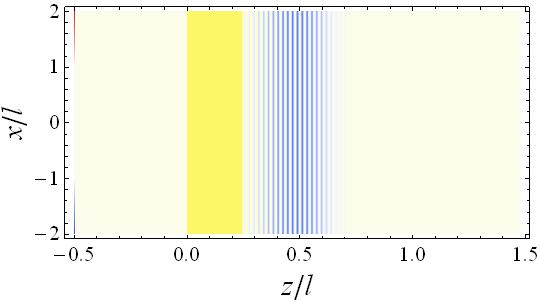}\includegraphics[width=.63cm]{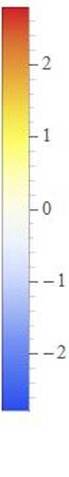}
\caption{Normalized charge density plot $1\!-\!n_e/n_0$ under the same conditions as in fig. \ref{graphs}
after about 25 fs (up) and 70 fs (down),  
i.e. resp. before and after the maximal penetration time $\bar t(0)\!=\!51$fs; in the latter picture the electrons travelling backwards
make  light yellow-blue striped the region between the yellow  and the white-blue striped ones.}
\label{chargedensityplot1}
\end{figure}

As a simplest illustration of the approach, and for later application to a step-shaped initial density, we first consider the case that \ $ \widetilde{n_0}( Z ) =n_0$. \ Then $F_e^z$ is
 the force  of a harmonic oscillator (with equilibrium at $z_e\!=\!Z$)
$F_e^z(z_e,\!Z)\!=\!- 4\pi n_0 e^2  [z_e\!-\! Z]\!=\!- 4\pi n_0 e^2 \Delta$; \ the $Z$-dependence disappears completely in
(\ref{heq1}-\ref{heq2}), which  reduces to the auxiliary Cauchy problem
\bea
&& \Delta'=\displaystyle\frac {1\!+\!v}{2s^2}\!-\!\frac 12,\quad
 s'=M\Delta,\qquad \Delta(0)\!=\!0, \:\:   s(0)\!=\! 1,\qquad\label{e1}
\eea
where $ M \!\equiv\!4\pi e^2n_0/mc^2$. \ The potential energy in (\ref{hamiltonian}) takes the form 
$\U(\Delta,Z)\!\equiv\!M\Delta^2/2$.
Problem  (\ref{e1}), and hence also its solution $\big(\Delta(\xi),s(\xi)\big)$, the value of the energy
as a function of $\xi$ and the functions
 defined in (\ref{defYXi}), are $Z$-independent. It follows  $\partial_Z\hat\Delta\!\equiv\!0$ and by (\ref{invZtoz})
the automatic invertibility of $z_e(t,\!Z)$;
moreover, the inverse function $Z_e(t,\!z)$ has the closed form  
\be
 Z_e(t,z)=ct\!\!-\!\Xi(ct\!-\!z)=z\!\!-\!\Delta(ct\!-\!z)                                                \label{sol'}
\ee 
[here $\Xi(\xi)\!\equiv\!\xi\!+\!\Delta(\xi)$], what makes the  solutions (\ref{sol}) of the system of functional equations  (\ref{functeq}), 
as well as those of (\ref{lageul}), completely explicit in terms of $\Xi$ and the inverse $\Xi^{-1}$ only.
As a consequence, all Eulerian fields depend on $t,z$ only through $ct\!-\!z$ (i.e. evolve as travelling-waves).
In fig.  \ref{pqCyclesM26}-left we plot some solution of (\ref{e1}).
If $v(\xi)\!\equiv\!v_c\!\equiv$const  
all paths $P(\xi;\!Z)$ are cycles  around  $C$ (fig. \ref{pqCyclesM26}-right), corresponding to periodic solutions.
Within the bulk  electron trajectories  for slowly modulated laser pulse like the ones considered in section \ref{experiments} are tipically as plotted in fig. \ref{Traj2};  in average they have no transverse drift, but  a longitudinal forward/backward one.
Fig. \ref{chargedensityplot1} shows a couple of corresponding  charge density plots.

\section{3-dimensional effects}
\label{3D-effects}

We now discuss the effects of the finiteness of $R$. 
For brevity, for any nonnegative $r,L$ we shall denote as $C_r$ the infinite cylinder
of equation $\rho\!\le\! r$, as $C_r^{\scriptscriptstyle L}$ the cylinder of equations $\rho\!\le\! r$, $0\!\le\!z\!\le\! L$.
The ponderomotive force of the pulse will boost forward (as in fig. \ref{Traj2}) only the small-$Z$ electrons located  within  
(or nearby) $\RC$. These Forward Boosted Electrons (FBE) will be thus completely expelled
out of a cylinder which will reach its maximal extension $\RC^{\:\zeta}$ around the time $\bar t(0)$ 
of maximal longitudinal penetration  $\zeta\!\equiv\!\hat\Delta[\bar \xi(0),\!0]$ of the  $Z\!=\!0$ electrons.
The displaced charges modify $\bE$. 
By causality (see appendix \ref{Rcond}), for $\bx$ near the $\vec{z}$ axis $\bE(t,\bx)$ 
 is the same as in  the plane wave case for $t\!\lesssim\! \bar t(0)\!+\!R/c$, 
and smaller afterwards. We choose $\widetilde{n_0},R$ so that 
they fulfill 
\be
 \frac{[t_{ex} \!-\!\bar t]c}R \sim 1,   \quad
r \equiv R-\frac{\zeta(t_{ex}\!-\!l/c)}{2(t_{ex}\!-\!\bar t)}\: \theta(ct_{ex}-l ) > 0                                   \label{req}
\ee
and condition (\ref{condapprox})  for  all $x\!=\!(t,\bx)$ such that $t\!\lesssim\! \bar t(0)\!+\!R/c$; 
here $\bar t\!\equiv\!\bar t(0)$, $t_{ex}\!\equiv\!t_{ex}(0)$ are
the times of maximal penetration and of
expulsion from the bulk of the $Z\!=\!0$ electrons [see (\ref{defbart_tex})].
[As $\widetilde{n_0}$ grows from zero the right-hand side of (\ref{condapprox}) does as well, whereas 
$\bar t(0),t_{ex}(0)\!-\!\bar t(0)$ decrease].
In appendix \ref{Rcond} we show that conditions (\ref{req}) respectively ensure  that these FBE, 
at least  within an inner cylinder $\rho\!\le\! r\!\le\! R$: i)
move approximately as in section \ref{Planewavessetup} until their expulsion; ii) are expelled before
Lateral Electrons (LE), which are initially located outside the surface of 
$\RC$ and are attracted towards the $\vec{z}$-axis (see fig. \ref{plasma-laser2}.3), obstruct their way out.
For the validity of our model we must  a posteriori check also that the expelled
$C_r$ electrons remain in $\RC$,
\be
\mbox{ i.e. their transverse oscillations $\Delta x_e$ are  $\ll R$}.
\label{condapprox'} 
\ee

\begin{figure}
\includegraphics[width=8cm]{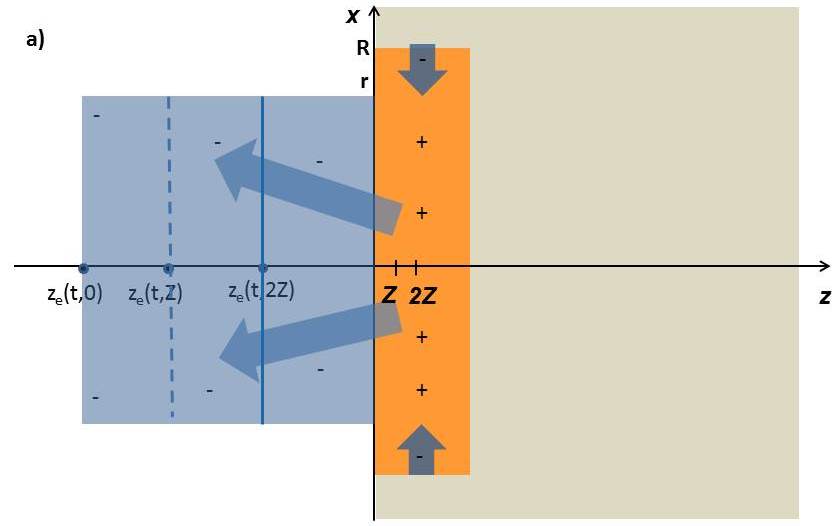}
\includegraphics[width=8cm]{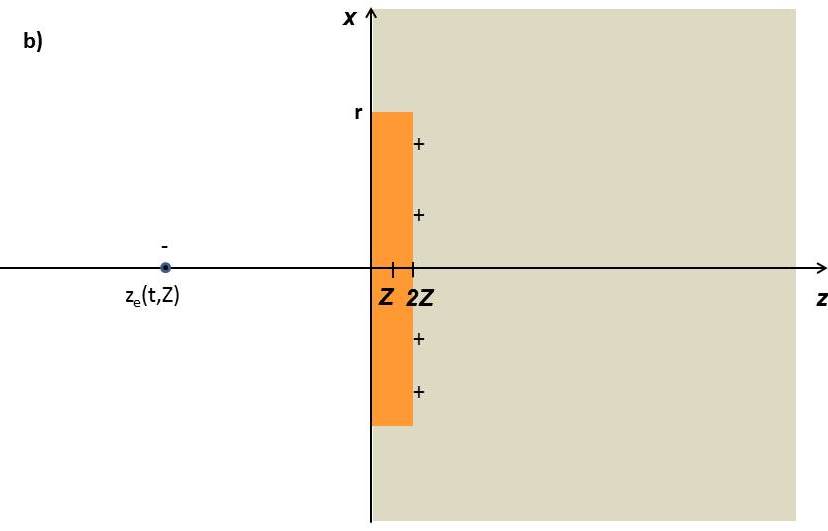}
\caption{a) schematic picture of the expected charge
distribution shortly after the expulsion  (long arrows) of surface electrons; short arrows 
represent the inward motion of the lateral electrons; b) simplified charge distribution 
generating the effective potential energy.
}
\label{pancake}
\end{figure}

\begin{figure*}
\includegraphics[width=8cm]{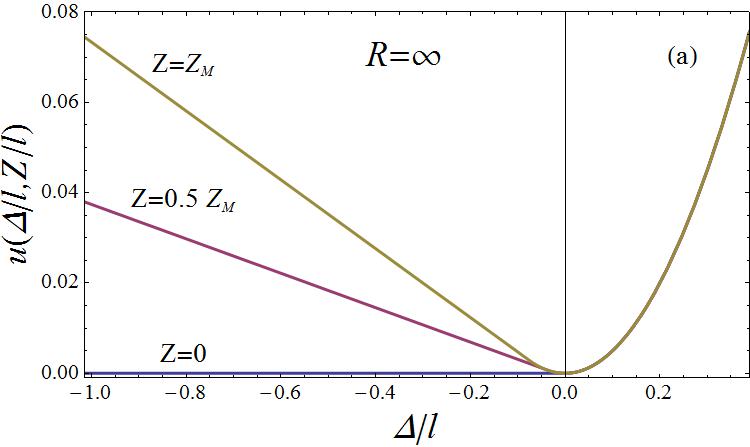}\hfill
\includegraphics[width=8cm]{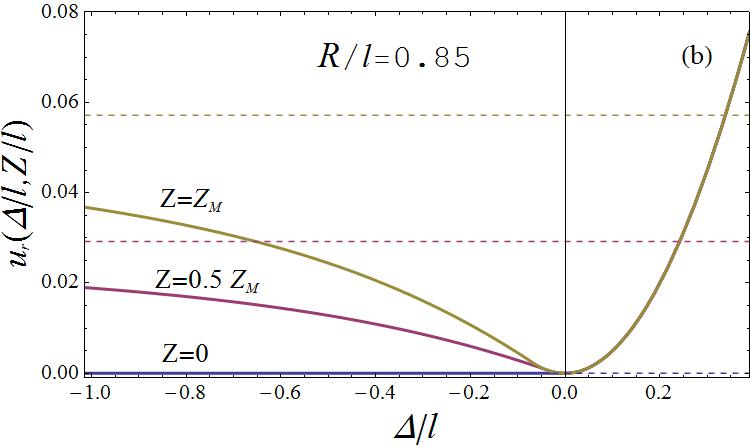}
\caption{Rescaled longitudinal electric potential energies $u\!\equiv\! \U/4\pi n_0e^2\!l^2$,
$\ru\!\equiv\! \rU /4\pi n_0e^2\!l^2$ for
(left) idealized plane wave $R/l\!=\!\infty$ or (right) for $R/l\!=\!0.85$,
plotted as functions of $\Delta$ for
$Z/\ZM=0,.2,.4,.6,.8,1$; the horizontal dashed lines are the left asymptotes of $\ru$ for
the same values of $Z/\ZM$.  Here the initial electron density is step-shaped: $\widetilde{n_{0}}(\!Z\!)\!=\!n_0\theta(\!Z\!)$.}
\label{PlotElForce}
\end{figure*}

In the plane model the expelled $Z\!>\!0$ electrons cannot escape to $z\to\!-\infty$ because
 are decelerated by the constant electric force $\widetilde{F_e^z}(t,Z)\!=\!4\pi e^2\widetilde{N}(Z)\!>\!0$,
see (\ref{definitions}). 
The real electric force $\widetilde{F_e^{{\scriptscriptstyle zr}}}\!>\!0$ acting on the $C_r$ electrons
after expulsion is generated by charges  localized in $\RC$; hence
 $\widetilde{F_e^{{\scriptscriptstyle zr}}}\!\propto \!1/z_e^2$ as 
$z_e \!\to\! -\infty$,  and the escape of expelled electrons is no more excluded {\it a priori}.
Moreover, since $\widetilde{F_e^z}(t,\!0)\!=\!0$, it should be also
$\widetilde{F_e^{{\scriptscriptstyle zr}}}(t,\!0)\!=\!0$, allowing the escape of the $Z\!=\!0$ electrons; 
 by continuity there will exist  some positive $Z\!_{{\scriptscriptstyle M}}\!\le\!Z_b$ such 
that the $C_r^{\ZM}$ electrons  escape to infinity. 
We stick to  estimate $\widetilde{F_e^{{\scriptscriptstyle zr}}}$
on  the $\vec{z}$-axis  electrons;
we assume that  after the pulse has overcome them, they move along the $\vec{z}$-axis.
Actually this will be justified below  if $\hat u^{{\scriptscriptstyle \perp}}(l)\!\simeq\! 0$, 
which in turn holds if, as usual,  $l\!\gg\!\lambda$ [see the comments after (\ref{finaldeviation})]. 
In fig. \ref{pancake} a) we schematically depict the charge distribution expected shortly after the expulsion. 
The light blue area is occupied only by the $\bX\!\in\!C_r^{Z\!_{{\scriptscriptstyle M}}}$  
electrons. The orange area is positively charged due to an excess of ions.
For any $Z$-electrons moving along
the $\vec{z}$-axis consider  the surfaces $S_0,S_1,S_2$ occupied at time $t$ by the 
$\bX'\!\in\! C_r$ electrons
respectively having
$Z'\!=\!0,Z,Z_2(Z)$, where $Z_2(Z)$ is defined by the condition $\widetilde{N}(Z_2)\!=\!2\widetilde{N}(Z)$,
which ensures that the electron charges contained between $S_0,S_1$ and $S_1,S_2$ are equal
(in the figure $S_0,S_1,S_2$ are respectively represented by the  left border of the
blue area, the dashed line and the solid line).
The longitudinal electric force $\widetilde{F_e^{{\scriptscriptstyle zr}}}$ 
acting  at time $t$ on this $Z$-electron is nonnegative and
can be decomposed and bound as follows \cite{FioFedDeA14}:
$$
0\le \widetilde{F_e^{{\scriptscriptstyle zr}}}\!(t,\!Z)
=-e\widetilde{E_{{\scriptscriptstyle -}}^{{\scriptscriptstyle z}}}\!(t,\!Z)\!-\!e
\widetilde{E_{{\scriptscriptstyle +}}^{{\scriptscriptstyle z}}}\!(t,\!Z)\le \rF\![\widetilde{\Delta}(t,\!Z),\!Z].
$$
Here $\widetilde{E_{{\scriptscriptstyle -}}^{{\scriptscriptstyle z}}}(t,Z)$ stands for the part of the
longitudinal electric field generated by the electrons between $S_0,S_2$;
since those between $S_0,S_1$ have by construction the same charge as those 
between $S_1,S_2$, but are more dispersed, it will be
$-e\widetilde{E_{{\scriptscriptstyle -}}^{{\scriptscriptstyle z}}}(t,Z)\le 0$. The part
$-e\widetilde{E_+^{{\scriptscriptstyle z}}}(t,Z)$  of $\widetilde{F_e^{{\scriptscriptstyle zr}}}$ 
 generated by the ions and  the remaining electrons (at the right of $S_2$)
 will be smaller than the 
force $\rF$ generated
by the charge distribution of fig. \ref{pancake} b), where  the remaining
electrons are located  farther from $(0,0,z_e)$ (in their initial positions $\bX'$,
not in the ones at $t$) and hence generate a smaller repulsive force. This explains the second
inequality in the equation.
In appendix \ref{Renergies} we show that  for \ $z_e\!\equiv\!Z\!+\!\Delta\!\le\!0$
\bea
 \frac{\rF(\Delta,\!Z)}{2\pi e^2 }
=2\widetilde{N}\!(\!Z\!)-\!\!\int_0^{Z_2(Z)}\!\!\!\!\!\!\!\!\!dZ' \frac { \widetilde{n_0}(Z') (Z'\!-\!z_e)}{
\sqrt{\! (Z'\!\!-\!z_e)^2\!+\!r^2}}. \qquad\label{Fzer}
\eea
Commendably, $\rF$ is conservative, nonnegative and goes to zero as $\Delta\!\to\!-\infty$, 
while it reduces to zero for $Z\!=\!0$ and to
$4\pi e^2\widetilde{N}\!(\!Z\!)$ as $r\!\to\!\infty$, as  $\widetilde{F_e^z}$ in (\ref{definitions})$_3$; it
 becomes a function of $t$ (resp. $\xi$) through $\widetilde{\Delta}(t,Z)$  [resp. 
$\hat \Delta(\xi,\!Z)$] only. We therefore modify the dynamics outside the bulk replacing 
$F_e^z$ by $\rF$, or equivalently
$\U$ by $\rU$ in (\ref{hamiltonian}), where $\rU$ is continuous and equals $\U$ for 
$z_e\!\equiv\!Z\!+\!\Delta\!\ge\!0$, and 
the  potential energy (\ref{rU}) associated to $\rF$
for $z_e\!\equiv\!Z\!+\!\Delta\!\le\!0$; there $\rU$
is a decreasing function of $\Delta$ with finite left 
asymptotes  (\ref{asymptotes}). We will thus {\it underestimate} the final energy of the electrons, 
because $\rF$ is larger than the real electric force $\widetilde{F_e^{{\scriptscriptstyle zr}}}$
decelerating the electrons outside the bulk; this  makes our estimates safer.
In fig.  \ref{PlotElForce} we plot suitably rescaled $\U$  and $\rU$
for  $ \widetilde{n_0}( Z ) =n_0\theta(Z)$.
After the pulse is passed we can compute $\gamma$   as a  function of $\Delta,Z$ using energy conservation
$m{}c^2\gamma\!+\!\rU(\Delta,\!Z)\!=$const. 
For the expelled electrons the final relativistic factor $\rga\!(Z) \!\equiv \! \gamma_e(\Delta\!=\!-\infty,\!Z)$ 
 is the decreasing function (\ref{genrga'}).
  The maximum of $\rga\!(Z)$ is  $\gamma_{{\scriptscriptstyle M}}\!\equiv\!\rga\!(0)$.
  Let $\ZM\!\le\! Z_b$   be the $Z$ fulfilling $\rga\!(Z)\!=\!1$.
The estimated total number $N_e$,  electric charge (in absolute value) $Q$, and kinetic energy ${\sf E}$ of the \
$\bX\in\!C_r^{\ZM}$ \ escaped 
electrons are thus
\be
\ba{l}
N_e\!\sim\! \pi r^2 \widetilde N\!\left(\ZM\!\right),\qquad
Q\!\sim\!  e N_e,\\[6pt]
{\sf E}\!\sim\! m\pi c^2r^2 \!\!\! \displaystyle\int_0^{\ZM}\!\!dZ\, \widetilde{n_0}(Z)
[\rga\!(Z)\!-\!1]. 
\ea                                 \label{NQE}
\ee
The number of escaped
$\bX'\!\in\!C_r^{\ZM}$ 
electrons with 
$Z\!\le\! Z'\!\le\!Z\!+\!dZ$ is 
estimated as $\pi r^2 \widetilde{n_0}(Z)dZ$, 
that with relativistic factor
between $\gamma$ and $\gamma\!+\!d\gamma$ is estimated as
$dN\!=\!\pi r^2 [ \widetilde{n_0}(Z)/|d \rga\!/dZ|]_{Z=\hat Z(\gamma)}\,d\gamma$, \
where $\hat Z(\gamma)$ is the inverse of $\rga(Z)$ (a strictly decreasing  function, see appendix \ref{Renergies}).
Hence the  fraction of escaped electrons with final relativistic factor
between $\gamma$ and $\gamma\!+\!d\gamma$ is
 estimated as $\nu(\gamma)d\gamma$, where 
\be
\nu(\gamma)\!\equiv\!\frac 1{N_e}\frac{dN}{d\gamma}
=\frac 1{ \widetilde N\!\!\left(\ZM\right)}
 \left.\frac{ \widetilde{n_0}(Z)}{\left|d \rga\!/dZ\right|}\right|_{Z=\hat Z(\gamma)}          \label{genspectrum}
\ee
determines the associated energy spectrum.  
As $\alpha^{{\scriptscriptstyle \perp}}(\xi)\!=\!\alpha^{{\scriptscriptstyle \perp}}(l)\!$ if $\xi\!\ge\! l$, by (\ref{u_es_e})
 the final transverse deviation of the escaped electrons will be
\be
\frac{\beta_f^{{\scriptscriptstyle \perp}}}{\beta_f^z}(Z)
\!=\!\frac{u_f^{{\scriptscriptstyle \perp}}}{u_f^z}(Z)
\!=\!\frac{u_f^{{\scriptscriptstyle \perp}}}{\sqrt{\gamma_f^2(Z)\!-\!1\!-\!u_f^{{\scriptscriptstyle \perp}2}}},    
  \label{finaldeviation}
\ee
where $u_f^{{\scriptscriptstyle \perp}}\equiv \hat u^{{\scriptscriptstyle \perp}}\!(l)$. This is an increasing function of $Z$, because  $\gamma_f(Z)$ is decreasing.
 If \ $\lambda\!\ll\! l$ then $ u_f^{{\scriptscriptstyle \perp}}\!\simeq\!0$ \ (see next section), and  (\ref{finaldeviation}) is
negligible unless \ $Z\!\simeq\!\ZM$.

\subsection{Step-shaped initial density}

\begin{figure}
\includegraphics[width=7.3cm]{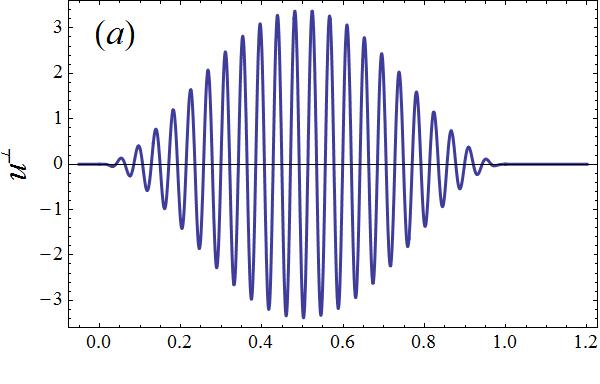}
\includegraphics[width=7.3cm]{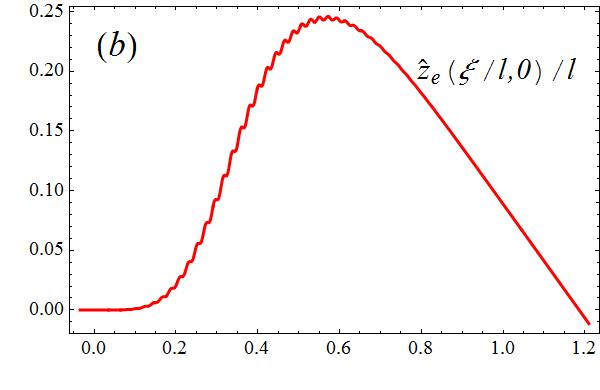} 
\includegraphics[width=7.3cm]{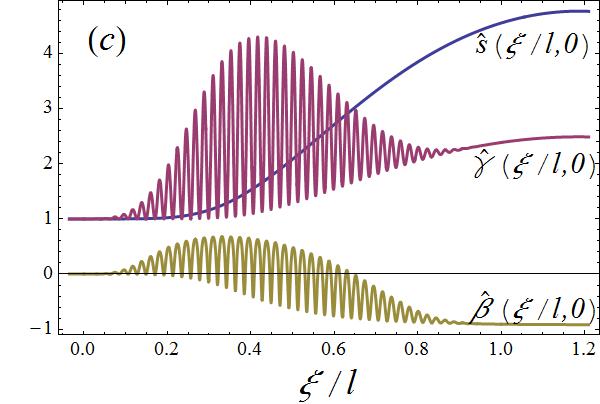} 
\caption{(a) Laser pulse  of average intensity $I\!=\!10^{19}$W/cm$^2$ and shape  as in section \ref{experiments},
with $l\!=\!18.75\mu$m.
(b-c)  Corresponding solution of (\ref{heq1}-\ref{heq2}) for  initial density $ \widetilde{n_0}(Z)\!=\!n_0\theta(Z)$,
with  $n_0\!=\! 21 \times 10^{17}$cm$^{-3}$   (i.e. $Ml^2\!=\!26$).
}
\label{graphs}
\end{figure}

\begin{figure*}
\includegraphics[width=17cm]{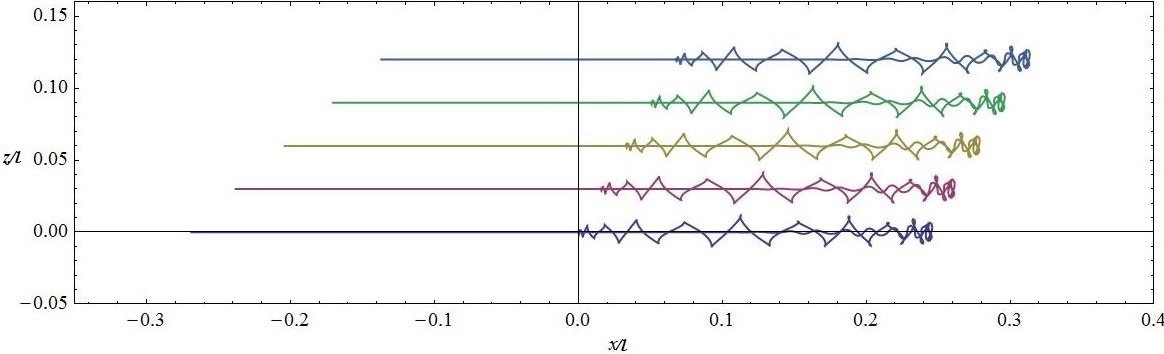} 
\caption{Trajectories gone in ca. 150 fs by  electrons initially located at 
 $Z/\ZM=0,  0.25 , 0.5 ,  0.75 ,1$ under conditions as in fig. \ref{graphs}.  
}
\label{Traj2}
\end{figure*}

If $ \widetilde{n_0}( Z ) \!=\!n_0\theta(Z)$  then   $\widetilde{N}( Z ) \!=\!n_0\theta(Z)Z$, \ and   
for $Z\!\ge\! 0$ 
\bea
\frac{\rF(\Delta,\! Z)}{2\pi n_0e^2}
\!=\! \left\{\!\!\ba{ll}
- 2 \, \Delta \quad  \mbox{(elastic force)},\: &z_e\!>\!0,\\[4pt]
 \!2Z\!+\!\sqrt{\!\! (\!Z\!\!+\!\!\Delta\!)^2\!\!+\!r^2}\!-\!\sqrt{\!\!  (\!Z\!\!-\!\!\Delta\!)^2\!\!+\!r^2},
\:\:   &z_e\!\le\!0.
\ea\right.\quad           \label{conservativeF}
\eea
Since the first expression is the same as in the case  $ \widetilde{n_0}( Z ) \!=\!n_0$, the motion of the $Z$-electron
will be as in subsection \ref{Auxiliary} until $\xi\!=\!\xi_{ex}(Z)$.
The second expression goes to  the constant force $4\pi n_0 e^2 Z$ as $r\!\to\!\infty$, as expected.
The motion for $\xi\!>\!\xi_{ex}(Z)$ will be studied in detail in \cite{Fio16}; we plot the  graphs
of a typical solution (until  the expulsion)  in fig. \ref{graphs} and 
 a few corresponding electron trajectories in Fig. \ref{Traj2}. 
We can readily understand that it will be
$\partial_{{\scriptscriptstyle Z}}\hat z_e(\xi,\!Z)\!\!>\!0$ for all $\xi$ and $0\!\le\!Z\!\le\! \ZM$,
since this holds for $\xi\!\le\!\xi_{ex}(Z)$ [by the comments following (\ref{e1})], and both
$\xi_{ex}(Z)$ and the decelerating force $\rF(\Delta,\! Z)$ (outside the bulk) increase with $Z$, while the 
speed of exit from the bulk decreases with $Z$,
whence the distance between electrons with different $Z$  increases with $\xi,t$.
The $Z_b$ introduced before (\ref{defxiex}) is now the solution of the equation 
$\sqrt{1\!+\!v(l)}\!+\!MZ_b^2/2\!=\!h$, i.e. the $Z$ corresponding to the zero longitudinal velocity and the final
value of the energy $h$ after the interaction of the pulse;  one can determine $h$ evaluating $H$
at $\xi\!=\!l$,  $h\!=\!\frac 12\{s(l)\!+\![1\!+\!v(l)]/s(l)\!+\!M[\Delta(l)]^2\}$. Hence,
\be
Z_b=\sqrt{[\Delta(l)]^2+\left[s(l)\!-\!\sqrt{1\!+\!v(l)}\right]^2\!/2M s(l)}.
\ee
$\rga(Z)$, $\nu(\gamma)$ admit rather explicit forms (\ref{rga}),  (\ref{spectrum}).
In  section \ref{experiments} we plot spectra $\nu(\gamma)$ corresponding to several $n_0$ and intensities.
Moreover, \ $Q\!=\!\pi r^2 e n_0\ZM$, \ ${\sf E}\!=\!
\pi r^2  n_0mc^2 \!\!\int_0^{\ZM}\!\!dZ[\rga\!(Z)\!-\!1]$.
Finally,  if $\xi_{ex}(0)\!<\!l$ then $\delta \bu\!^{{\scriptscriptstyle\perp}}$ in (\ref{condapprox}) becomes 
\be
\delta \bu\!^{{\scriptscriptstyle\perp}}\!(t,z)\!=\!\frac{M}{2}\!\! 
\displaystyle\int\limits^{ct-z}_{0}\!\!\!\!\! d\xi'\,
\frac{\hat\bu\!^{{\scriptscriptstyle\perp}}\!(\xi')}{s(\xi')}\!\!\displaystyle\int\limits^{ct+z}_{0}\!\!\!\!d\xi'' \,\theta\!\left[\frac{\xi'\!\! \!+\!\xi''}2\!-\!\Xi(\xi')\right]\!\! .
\ee

\section{Numerical results}
\label{experiments}

We assume for simplicity that the pulse is a 
slowly modulated sinusoidal function  linearly polarized in the $x$ direction: \
$\Be^{{\scriptscriptstyle \perp}}(\xi)=\epsilon_s(\xi){\hat\bx}\cos k\xi$, \ 
the modulating amplitude $\epsilon_s(\xi)\!\ge\! 0$ is nonzero only for  $0\!<\!\xi\!<\!l$, 
and slowly varies on the scale of the period $\lambda\equiv\!2\pi/k \!\ll\! l$, i.e.  $\lambda |\epsilon_s'|\!\ll\!|\epsilon_s|$ on the support of $\epsilon_s$. Integrating by parts  we find
 \ $\Ba^{{\scriptscriptstyle \perp}}(\xi)\!=\! 
{\hat\bx}\epsilon_s(\xi)(\sin k\xi)/k\!+\!O(1/k^2)$ \cite{Fio16c} and, in terms of the rescaled amplitude
$w(\xi)\!\equiv\! e\epsilon_s(\xi)/kmc^2$,
\bea
\hat\bu^{{\scriptscriptstyle \perp}}(\xi)\!\simeq\! {\hat\bx}\,w(\xi) \,\sin(k\xi),\qquad
v(\xi)\!\simeq\! w^2(\xi) \,\sin^2(k\xi) ,   \qquad                   \label{vapprox}    
\eea    
where \ $a\!\simeq\!b$ \ means \ $a\!=\!b\!+\!O(1/k^2)$. Note that, as $\epsilon_s(\xi)\!=\!0$ for $\xi\!\ge\! l$,
this implies  $ u_f^{{\scriptscriptstyle \perp}}\!=\!  \hat u^{{\scriptscriptstyle \perp}}\!(l)\!\simeq\!0$, \  as anticipated.

If we approximate as $\chi\!_{{\scriptscriptstyle R}}(\rho)\!\equiv\!\theta(R\!-\!\rho)$
 the cutoff function in  (\ref{pump}), the average pulse intensity on
its support is \ $I\!=\! c\,\E\!/\pi R^2 l$.   
Here $\E$ is the EM energy carried by the pulse, 
\bea
\E=\!\int_V\!\!\!\!dV\frac{\bE^{{\scriptscriptstyle\perp}2}\!\!+\!\bB^{{\scriptscriptstyle\perp}2}}{8\pi}
\!\simeq\!\frac{R^2}{4}\!\! \int_0^l\!\!\!\!d\xi\,\Be^{{\scriptscriptstyle \perp}2}\!(\xi)
\!\simeq\!\frac{R^2}{8}\!\! \int_0^l\!\!\!\!d\xi\,\epsilon_s^2\!(\xi).
\qquad \label{pulseEn}
\eea
High power lasers  produce pulses where $\lambda\!\sim\! 1\mu$m and $\epsilon_s$ 
is  approximately gaussian, 
$\epsilon_s(\xi)\!\propto\! \exp\!\left[-(\xi\!-\!\xi_0)^2/2\sigma\right]$; 
$\sigma$ is related to the fwhm (full width  at half maximum) $l'$  of $\epsilon_s^2$
by $ \sigma\!=\!l'{}^2/4\ln2$. 
If initially matter is composed of atoms then
$\epsilon_s(ct\!-\!z)$ can be considered zero where it is under the ionization threshold, because the pulse 
has not converted matter into a plasma yet. Hence we  adopt 
as a modulating amplitude $\epsilon_s(\xi)$  the cut-off Gaussian
\bea
\epsilon_{g}(\xi)\!=\!b_g \exp\!\left[\frac{-(\xi\!-\!l/2)^2}{2\sigma}\right]
\theta(\xi)\theta(l\!-\!\xi) , \quad \sigma\!=\!\frac{l'{}^2}{4\ln2}, \qquad \label{w_g}\\
b_g^2\!=\!\frac{16\sqrt{\ln 2}}{\sqrt{\pi}} \frac{\E }
{R^2l'},\quad l^2=\frac{l'{}^2}{\sqrt{\ln 2}}\ln\!\left[\frac{\sqrt{\ln 2}\,mc^2\, \E\, (e\lambda)^2}{U_i\,\sqrt{\pi}\,l'\,(\pi Rm{}c^2)^2}\right]\!,\nonumber
\eea
where $U_i$  is the first ionization potential (for Helium $U_i\simeq 24eV$); the formula
for $b_g^2$ follows replacing the Ansatz (\ref{w_g})$_1$   in (\ref{pulseEn})
[neglecting the tails left out by the cutoff \ $\theta(\xi)\theta(l\!-\!\xi)$].
Numerical computations are easier if we adopt \cite{Fio14JPA} as $\epsilon_s(\xi)$
the following cut-off polynomial:
\bea
\epsilon_p(\xi)\!\equiv\!
\frac{b_p}4\!\left[\!1\!-\!\left(2\xi/l_p\!-\!1\!\right)^2\! \right]^2
\!\!\theta(\xi)\theta(l_p\!-\!\xi), 
\label{w_p}
\eea
$b_p,l_p$ are determined by the requirement to lead to the same fwhm and $\E$: \
$ b_p^2 \!=\!5040\,\E /R^2l_p$ and $l_p\!=\!5l'/2$. 

\medskip
We now present the results of extensive numerical simulations based on the
experimental parameters available already now at the FLAME facility \cite{GizEtAl13} or 
in the near future at the ILIL facility \cite{footnote4}:
\ $l'\!\simeq\! 7.5 \mu$m (implying $l_p\!=\!18.75\mu$m), \ $\lambda\!\simeq\! 0.8 \mu$m 
(implying $kl_p\!=\!2\pi l_p/\lambda\!\simeq\!147$), \
 \ $\E\!=\!5$J, \   \ and 
 $R$ tunable by focalization in the range \ $10^{-4}\div 1$ cm. 
We model the electron density:
first as the step-shaped one \  $ \widetilde{n_0}(Z)\!=\!n_0\theta(Z)$ 
(this allows analytical derivation of more results); \
then as a function  smoothly increasing from zero to the asymptotic value $n_0$,
 with substantial variation in the interval  \ $0\!\le\!Z\!\le\! L\!\equiv\!20\mu$m \
(as motivated by experiments, see section \ref{Final}),
more precisely  \ $ \widetilde{n_0}(Z)\!=\!n_0\,\theta(Z)\tanh(Z/L)$.
We have numerically solved the corresponding systems (\ref{heq1}-\ref{heq2}) and proceeded as in
section \ref{3D-effects},
for  $R\!=\!16,15,8,4,2,1\mu$m [resp. leading to average intensities $I/10^{19}$(W/cm$^2)\!\simeq\!1,1.1,4,16,64,255$], 
$n_0$ in the range $10^{17}$cm$^{-3}\!\le\! n_0\!\le\! 3\times 10^{20}$cm$^{-3}$ and $Z\!\le\!\ZM$; all results follow from these solutions.

\begin{figure*}
\includegraphics[width=7.8cm]{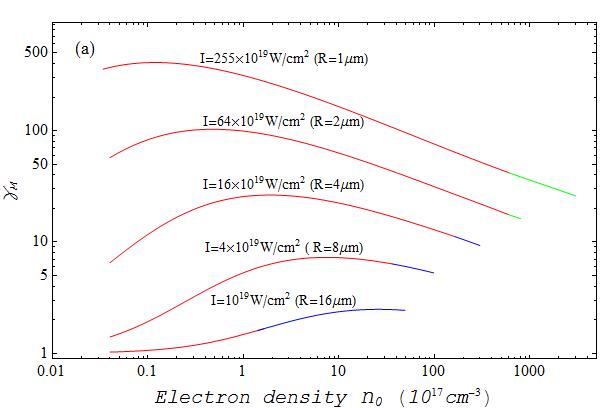} \includegraphics[width=4.55cm]{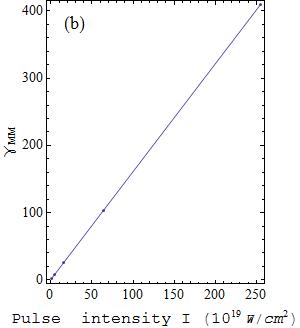}
\includegraphics[width=5cm]{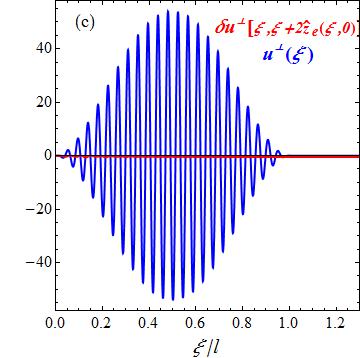}
\caption{Left: relativistic factor $\gamma_{{\scriptscriptstyle M}}$ of the  
$Z\!=\!0$ expelled electrons (the maximal one) as a function of the step-shaped initial electron density $n_0$,
for few values of the intensity $I$; the maximum of each graph is denoted as $\gamma_{{\scriptscriptstyle MM}}$. Center: $\gamma_{{\scriptscriptstyle MM}}$ vs. $I$. Right: $\bu\!^{{\scriptscriptstyle\perp}}$ \& its correction $\delta \bu\!^{{\scriptscriptstyle\perp}}$ along the $\bX\!\!=\!0$ electrons' worldlines 
for  $n_0\!=\!24\!\times\!10^{19}$cm$^{-3}$, 
$I\!=\!255\!\times\!10^{19}$W/cm$^2$: $\delta \bu\!^{{\scriptscriptstyle\perp}}$ is negligible.}
\label{EnTot_vs_n0}
\end{figure*}
\begin{figure*}
\includegraphics[width=5.9cm]{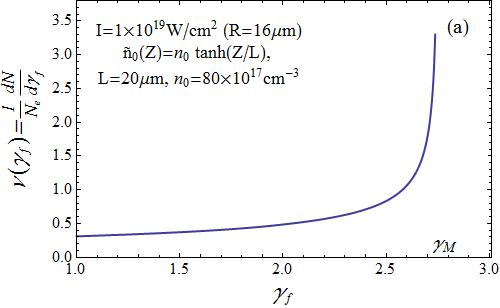} \hfill
\includegraphics[width=5.9cm]{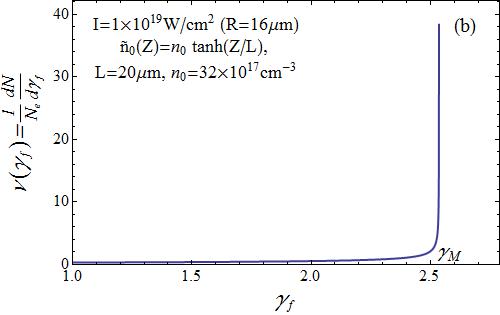} \hfill
\includegraphics[width=5.9cm]{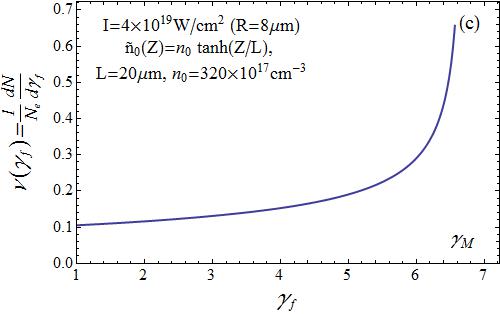}\\
\includegraphics[width=5.9cm]{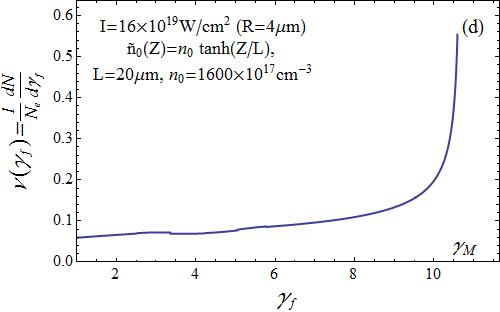} \hfill
\includegraphics[width=5.9cm]{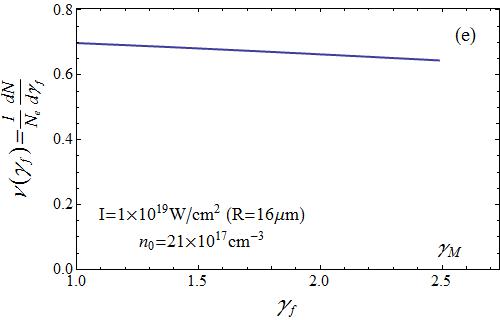} \hfill 
\includegraphics[width=5.9cm]{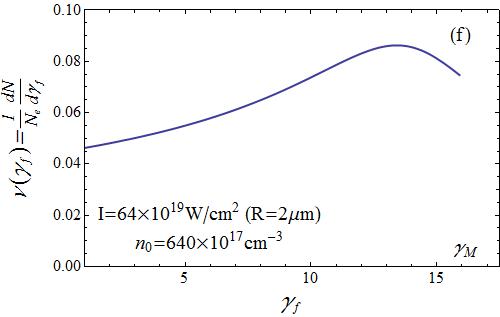}
\caption{Sample spectra of the  expelled electrons for pulse amplitudes of the form (\ref{w_p}) 
with continuous initial electron densities  \ $ \widetilde{n_0}(Z)\!\equiv\!n_0 \theta(Z)\tanh(Z/L)$, \  
 $L\!=\!20\mu$m (graphs a-d),  or step-shaped initial electron densities \ $ \widetilde{n_0}\!\equiv\!n_0 \theta(Z)$  
(graphs e-f). 
The values of $n_0$ and of the average pulse intensity $I$ are the same as in Table \ref{tab1}.}
\label{nu_vs_gamma}
\end{figure*}

In fig. \ref{EnTot_vs_n0}-left we plot the  maximal final relativistic factor 
$\gamma_{{\scriptscriptstyle M}}$ of the expelled electrons as a function of $n_0$, with the above values
of $I$ and $ \widetilde{n_0}(Z)\!=\!n_0\theta(Z)$;
each graph stops where $n_0$ becomes too large for conditions (\ref{condapprox}), (\ref{req})$_1$,  
or (\ref{condapprox'})  to be fulfilled and is red where condition (\ref{req})$_2$  is no more fulfilled.
The latter prevents  collisions with the LE and becomes superfluous if the 
target is a solid cylinder of radius $R$ (since then there are no LE) \cite{footnote5}; 
the $I\!=\!64,255\!\times\!10^{19}$W/cm$^2$ graphs are plot green for densities corresponding to
the lightest solids (aerogels) available today.
As expected \cite{FioFedDeA14}: 1)  as $n_0\!\to\!0$ \
$\gamma_{{\scriptscriptstyle M}}\!-\!1\!\propto\!n_0I^2$; 2)
each graph $\gamma_{{\scriptscriptstyle M}}(n_0;I)$ has a unique maximum
$\gamma_{{\scriptscriptstyle MM}}(I)\!\equiv\!\gamma_{{\scriptscriptstyle M}}(n_{0{\scriptscriptstyle M}};I)
$ at $n_{0{\scriptscriptstyle M}}\!\sim\!\bar n_0$, where $\bar n_0$ is the density
making $\bar\xi(0)\!=\!l/2$,  namely such that the $Z\!=\!0$ electrons
reach the maximal penetration \  $\zeta\!=\!\hat\Delta(l/2,0)$
 when they are reached by the pulse maximum. The dependence  of $\gamma_{{\scriptscriptstyle M}}$
on $n_0$ is anyway rather slow.
The striking $\gamma_{{\scriptscriptstyle MM}}(I)\propto I$ behaviour shown in fig. \ref{EnTot_vs_n0} up-center
hints at scaling laws and will be discussed elsewhere. 
In figures \ref{nu_vs_gamma} we plot sample spectra $\nu(\gamma)$ for 
$I/10^{19}$(W/cm$^2)\!\simeq\! 1,4,16,64$ and 
$\widetilde{n_0}$ compatible with  (\ref{condapprox}), (\ref{req}),  (\ref{condapprox'}).
In table \ref{tab1} we report our main predictions  for the same  $I$ (equivalently, $R$) and $\widetilde{n_0}$. 
The final energies of the expelled electrons  range from
few to about 15 MeV. The spectra (energy distributions) are rather flat for the  step-shaped densities,
albeit they become more peaked near $\gamma_{{\scriptscriptstyle M}}$ as $n_0$ grows; 
if  $ \widetilde{n_0}(Z)$ grows smoothly from zero to about the asymptotic value $n_0$ in the interval $0\!\le\! Z\!\le\! L\!\sim\! 20 \mu$m,
 they can be made much better (almost monochromatic) by tuning $L$.
The collimation of the expelled electron bunch is very good, by  (\ref{finaldeviation}); in all cases considered in table  \ref{tab1} we find deviations $\beta_f^{{\scriptscriptstyle \perp}}/\beta_f^z$ of $1\div 2$ milliradiants
for the  $(\rho,Z)\!=\!(0,0)$  and $4\div 10$ milliradiants
for the $(\rho,Z)\!=\!(0,0.9 \ZM)$ expelled electrons. 

\begin{table*}[ht] $\!\!\!\!\!\!\!\!\!\!\!\!\!\!\!\!\!\!\!\!\!\!\!\!\!\!\!\!\!\!\!\!\!\!\!\!$
\begin{tabular}{|c|}
\hline
pulse energy ${\cal E}\!\simeq\!5$J, wavelength $\lambda\!\simeq\!0.8\mu$m, 
 fwhm $l'\!\simeq\!7.5\mu$m, spot radius $R\!\simeq\!1\div 16\mu$m\\
\hline
\begin{tabular}{|l|c|c|c|c|c|c|c|c|c|c|c|c|c|c|c|c|c|c|c|c|c|c|c|}
\hline
 & p & {\tiny PoP14} &  p & {\tiny PoP14}  & p & g  &\!\!& cp & cg &  cp &  cg &   cp &  cg &  cp &  cg  \\[2pt]
pulse spot radius $R \, (\mu$m)  & 15 & 15 & 16 & 16  &  2 & 2  &\!\!& 16 & 16 & 16 & 16 & 8 & 8 &  4 & 4\\[2pt]
mean intensity $I\,$(10$^{19}$W/cm$^2$)  &1.1 &1.1 & 1 & 1  & 64 &  64 &\!\!& 1 & 1 & 1 & 1 & 4 & 4 &16  & 16   \\[2pt]
initial el. density $n_0(10^{18}$cm$^{-3}$) &.64  & .64  & .64 & .64 &  64 & 64  &\!\!& 3.2 & 3.2 & 8 & 8  & 32  & 32 & 160  & 160   \\[2pt]
 ratio \ $[t_{ex}(0)\!-\!\bar t(\0)]c/R\!\!$ & 0.8  &  & .75 &   & 1.3 & 1.9  &\!\!& 0.8 & 0.8 &  0.6  & 0.7 & 1.1 & 1.1 &  1.7  & 2.1   \\[2pt] 
 ratio \ $r/R\!\!$ & 0.6 &   & 0.7 &  & 1 & 1  &\!\!&  0.6 & 0.7 & 0.8  &   0.9 & 0.8 & 1&   1  & 1  \\[2pt] 
ratio \ $\Delta x_e^{{\scriptscriptstyle M}}/R$   & .02 &  & .02 &  &  .14 & .09 &\!\!& .02  & .19 & .02 & .17 & .05  & .04 & .1  & .06 \\[2pt]
\hline
maximal relativistic factor $\gamma_{{\scriptscriptstyle M}}$ & 2.5 & 1.83 &2.3 &  1.65  & 16 & 14  &\!\!& 2.5 & 2.4 & 2.7  & 2.6  & 6.6  & 5.6 & 11  & 8.1 \\[2pt]
max. expulsion energy $H$(MeV)   & 1.3 &0.94 &1.2 & 0.9  &  8.1 & 7.2   &\!\!&  1.3 & 1.2 & 1.4   & 1.3   & 3.4  & 2.9 & 5.4  & 4.2  \\[2pt]
tot expelled  charge $|Q|(10^{-10}$C)  & 1.7 & 3.8 & 2.2 & 3.2 &  3.7 & 3.1    &\!\!& 1.9 & 2.2 & 3.7   & 2.9 & 3.5  & 4 & 3.8  & 3.5 \\[2pt]
tot. exp.  kin. energy ${\sf E}(10^{-4}$J)   & 0.7 &  & 0.7 & &  16 & 12   &\!\!& 0.8 & 0.8 & 1.6  & 1.1 & 4.5  & 4.0 & 8.6  & 5.8  \\[2pt]
\hline
\end{tabular}
\end{tabular}
\caption {Sample inputs and outputs for possible experiments. In the `p,g'   columns the initial electron densities are step-shaped, $ \widetilde{n_0}(Z)\!=\!n_0\,\theta(Z)$, and the amplitudes are resp. of the gaussian, 
polynomial forms (\ref{w_g}), (\ref{w_p}); in the `PoP14' columns we report  
results   computed in \cite{FioFedDeA14}  with poorer approximation.
   In the `cg,cp'   columns the initial electron densities are the continuous ones 
$\widetilde{n_0}(Z)\!=\!n_0\,\theta(Z)\tanh(Z/L)$ with $L\!=\!20\mu$m, and the amplitudes are resp. of the 
forms (\ref{w_g}), (\ref{w_p}).   
}
 \label{tab1}  
\end{table*}

We now discuss the conditions guaranteeing the validity of our model. 
The comments after (\ref{conservativeF}) show for all $\xi$ the invertibility 
of the maps $\hat z_e(\xi,\cdot):Z\!\mapsto\! z$
 in the interval $0\!\le\!Z\!\le\! \ZM$, and therefore the self-consistency of this 2-fluid MHD model,
in the step-shaped density case; numerical study of the map $\hat z_e(\cdot,\xi):Z\!\mapsto\! z$
 shows that this holds true also in the continuous density case.
 Numerical computations show that 
(\ref{condapprox}) is fulfilled at least on the $Z\!\le\!\ZM$ electrons' worldlines, 
 even with the highest densities considered here (see e.g. fig. \ref{EnTot_vs_n0} right).
Finally, the data in table \ref{tab1}
show that  (\ref{req}),  (\ref{condapprox'}) are fulfilled.

If we choose $\epsilon_s(\xi)$ as the cut-off gaussian, instead of the 
 cut-off polynomial, convergence of numerical computations is slower, 
but the outcomes do not differ significantly. Sample computations show
that choices of other continuous $\widetilde{n_0}(Z)$ 
lead to similar results, provided the function $ \widetilde{n_0}(Z)$ is increasing and 
significantly approaches the asymptotic value $n_0$ in the interval 
$0\!\le\! Z\!\le\! L\!\sim\! 20 \mu$m.

\section{Discussion, final remarks, conclusions}
\label{Final}

These results show that indeed the slingshot effect is a promising acceleration mechanism of electrons,
in that it extracts from the targets highly collimated bunches  of electrons with spectra which can be made peaked around the maximum energies by adjusting $R,\widetilde{n_0}$;
with laser pulses of a few joules and duration of few tens of femtoseconds (as available today in many laboratories) we find that the latter range up to about ten MeV (it would increase with more energetic pulses). 
The spectra (distributions of electrons as functions of the final relativistic factor $\gamma_f$),
 their dependence on the electron density and pulse intensity, the collimation and the backward direction of expulsion in principle allow to discriminate the slingshot effect from the LWFA or other 
acceleration mechanisms. In table  \ref{tab1} and fig. \ref{nu_vs_gamma}
 we have reported detailed  quantitative predictions of the main features of the effect for some possible choices of parameters in experiments at the present FLAME, 
the future upgraded ILIL facilities, or similar laboratories. Low density  gases  or aerogels (the lightest solids available today) are targets with appropriate electron densities. 

The steepest $z$-oriented density gradient  of a gas sample isolated in vacuum is attained just outside a nozzle expelling 
a supersonic gas jet in the $xy$ plane; across the lateral border of the jet  the density may vary from about zero 
to almost the asymptotic value $n_0$ in about $L\!\sim\!20\mu$m \cite{GizEtAl13}. Hence  if we choose
a supersonic helium jet as  the laser pulse  target the initial electron density
is reasonably approximated by the choice 
$\widetilde{n_0}(Z)\!=\!n_0\,\theta(Z)\tanh(Z/L)$, 
and the predictions of table  \ref{tab1}, fig. \ref{nu_vs_gamma} (a-d) are reliable. 
By the way, the values of $n_0$ considered in  table  \ref{tab1} are considerably higher than 
in typical LWFA experiments.

 Step-shaped $ \widetilde{n_0}(Z)$
 are unrealistic approximations of densities of gas samples, but reasonable ones of solids 
(for which $\Delta Z\!\ll\! \lambda$), provided  $n_0$ exceeds $48 \!\times\!10^{18}$cm$^{-3}$,
which is the electron density of  aerographene (the lightest aerogel so far: mass density=0.00016 g/cm$^3$). 
Silica areogels, with a wide range of densities from 0.7 to 0.001g/cm$^3$, electron densities of the order
of  10$^{20}$/cm$^{-3}$
  and porosity  from 50 nm  down to 2 nm in diameter (i.e. much smaller than $\lambda$) have been  produced and extensively studied \cite{PiePaj02,AmbBag06}.
Therefore the results of the last two `p,g' columns of table  \ref{tab1} 
[and the corresponding spectra, figure \ref{nu_vs_gamma} (e)] are applicable to aerogels, while those of the first four are presently only of academic interest.

The quantitative predictions of our model are based on a rather rigorous plane-wave, 2-fluid magnetohydrodynamic 
model \cite{footnote6} and simple,  but heuristic approximations for the 3D corrections,
which certainly affect their liability. We welcome numerical  3D simulations 
(particle-in-cell ones, etc.) to improve the latter. Experimental tests are easily feasible with 
the equipments presently available  in many laboratories.
We welcome experiments testing the effect.

\bigskip 
{\bf Acknowledgments.} 
We are pleased to thank R. Fedele, L. Gizzi for stimulating discussions.
G. F. acknowledges partial support by UniNA \& Compagnia di San Paolo under the grant  {\it STAR Program
2013}, and by COST Action MP1405  {\it Quantum Structure of Spacetime}.

\appendix
\section{Finite $R$ conditions}
\label{Rcond}

\begin{figure*}
\begin{center}
\includegraphics[width=6.2cm]{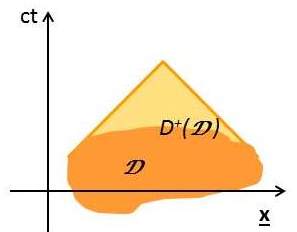}\hfill
\includegraphics[width=10.4cm]{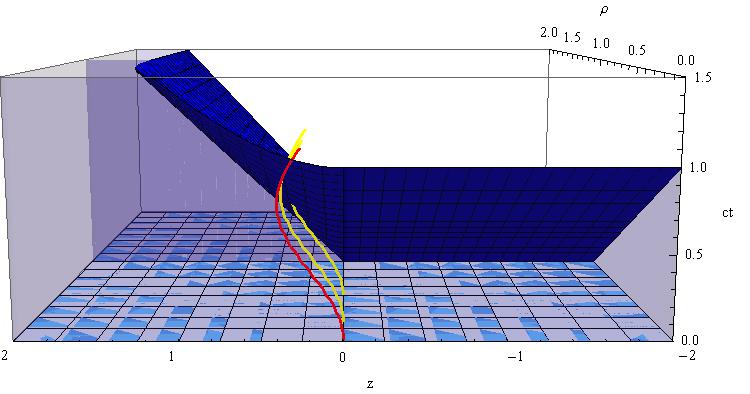}
\end{center}
\caption{Left: future Cauchy development $D^+(\D)$ of a generic domain $\D$. 
Right: $\D^0_{{\scriptscriptstyle 1}}$ (light blue) and $D^+(\D^0_{{\scriptscriptstyle 1}})$ (shaded
region between the blue and light blue hypersurfaces) in $(\rho,z,ct)$ coordinates 
(we have dropped the inessential angle $\varphi$);
worldlines of the  $\bX\!=\!0$ electrons (red) and of a couple of off-$\vec{z}$-axis electrons (yellow);
the former remain in $D^+(\D^0_{{\scriptscriptstyle 1}})$ longer.}
\label{causcomp}       
\end{figure*}

As known, for any spacetime region $\D$  its {\it future Cauchy development} $D^+(\D)$
is defined as the set of all points $x$ for which every past-directed causal (i.e. non-spacelike) 
line through $x$  intersects $\D$  (see fig. \ref{causcomp} left).
Causality  implies:
{\it  If two solutions of the system of  
dynamic equations coincide in some open 
spacetime region $\D$, then they coincide also in $D^+(\D)$.}
Therefore, knowledge of one solution determines also the other  (which we will distinguish by adding a prime to all fields)
in $D^+(\D)$. 

In the problem at hand
the solutions are exactly known for $t\!\le\!0$, i.e. before the laser-plasma interaction begins. 
We use  causality  adopting: {\bf 1.} as $\D$ a region $\D^0_{{\scriptscriptstyle R}}$ (see fig. \ref{causcomp} right)
 of equations  $-\epsilon\!\le\!t\!\le\! 0$ and either $\rho\!<\! R$ or $z\!>\!0$, with some $\epsilon\!>\!0$ (we can take also  $\epsilon\!=\!0$ if we assign on $\D^0_{{\scriptscriptstyle R}}$ also the time derivatives of the $A^\mu,\bu$); {\bf 2.}
as the known solution the plane one
induced (section \ref{Planewavessetup})  by the plane transverse electromagnetic 
potential, which  can be approximated 
as $\bA\!^{{\scriptscriptstyle\perp}}(t\!,z)\!=\!\Ba\!^{{\scriptscriptstyle\perp}}(ct\!-\!z)$ under the assumption
(\ref{condapprox});
{\bf 3.}  as the unknown solution  the ``real'' one induced by the ``real''  laser pulse $A^\mu_f(t,\!\bx)$, 
which we approximate as a potential leading to (\ref{pump}). It is easy to show that  $D^+(\D^0_{{\scriptscriptstyle R}})$
is the union of three regions, resp. of equations: {\bf a.} $z\!\ge\!ct\!\ge\!0$; {\bf b.}  $ct\!\ge\!z\!\ge\!0$ and 
$\rho\!+\!\sqrt{c^2t^2\!-\!z^2}\!\le\! R$; {\bf c.} $t\!\ge\!0$, $z\!<\!0$ and 
$\rho\!+\!ct\!\le\! R$  (see fig. \ref{causcomp}). In $D^+(\D^0_{{\scriptscriptstyle R}})$ the two solutions coincide,
in particular a ``real''  electron worldline $\bx_e'(t,\bX)$ remains equal to the plane solution
worldline $\bx_e(t,\bX)$ as long as $\bx_e(t,\bX)\!\in\!D^+(\D^0_{{\scriptscriptstyle R}})$. 

By continuity, we expect that the two solutions remain close to each other also
in a neighbourhood of  $D^+(\D^0_{{\scriptscriptstyle R}})$. This is confirmed
by estimates \cite{FioFedDeA14} involving the retarded electromagnetic potential
(in the Lorentz gauge $\partial\cdot A\!=\!0$) 
\be
 A^\mu\!(t,\!\bx)\!=\!A^\mu_f(t,\!\bx)\!+\!\!\int\!\! \!d^3x' \frac{j^\mu[t_r(t,\!\bx\!-\!\bx'),\bx']}{|\bx\!-\!\bx'|}, \qquad       
                     \label{retAmu}
\ee
i.e. the general solution of the Maxwell equation $\Box A^\mu\!=\!4\pi j^\mu$ 
with a current $ j^\mu(t,\bx)$ vanishing for $t\!<\!0$;
here $t_r(t,\!\bx\!-\!\bx')\!\equiv\!t\!-\! |\bx\!-\!\bx'|/c$, $A^\mu_f(t,\!\bx)$ fulfills $\Box A^\mu_f\!=\!0$ (determining the 
$t\!\to\!-\infty$ behaviour),
and  $\bE\!=\!\frac {-1}c\partial_t\bA\!-\!\!\nabla\!\! A^0\!$,  $\bB\!=\!\nabla\!\times\! \bA\!$. 
Since the formation of 
$\RC^{\:\zeta}$ is completed at $t\!=\!\bar t(0)$, and the `information' [encoded 
in (\ref{retAmu})] about the finite radius of $\RC^{\:\zeta}$
 takes a time $R/c$ to go from the lateral surface $\rho\!=\!R$ to the $\vec{z}$-axis,
then if eq. (\ref{req})$_1$ is fulfilled
 the  $\bX\!=\!0$ electrons (red worldline in fig. \ref{causcomp})  move approximately 
as in section \ref{Planewavessetup} until the expulsion.
Similarly, the $Z\!\simeq \!0$, $\rho\!\lesssim\!r$ electrons  (yellow worldlines in fig. \ref{causcomp})
move approximately as in section \ref{Planewavessetup} until $\bar t+ (R\!-\!r)/c$, i.e
get the main  backward boost  (acceleration is maximal
around $\bar t$). 
Eq. (\ref{req})$_2$ is equivalent to
\bea
\ba{llll}
 &t_{ex}\lesssim l/c; \qquad&\Rightarrow \quad  &r\simeq R; \\[4pt]
\mbox{or } &0\!<\! (t_{ex}\!-\!l/c)v^\rho_a \!<\! R \quad &\Rightarrow \quad & r\simeq R-(t_{ex}\!-\!l/c)v^\rho_a>0.
\ea                                      \nonumber 
\eea
If the left-hand side of the first line is fulfilled the surface electrons 
are expelled while the laser pulse is still entering the bulk
and thus producing an outward force that keeps the LE out of $\RC^{\:\zeta}$. 
Otherwise, the left-hand side of the second line ensures that the distance inward travelled by the
most dangerous LE (the $Z\!=\!0$ ones)  {\it after} the pulse has completely entered the bulk is less than $R$;
$v^\rho_a$ stands for the average $\rho$-component of the velocity of these LE. 
By geometric reasons  $v^\rho_a\!<\!v^z_a\!\equiv$ 
average $z$-component of the  $\bX\!=\!0$ electrons   velocity
in their backward trip within the bulk; 
our  rough estimate $v^\rho_a\!\simeq\! v^z_a/2\!=\!\zeta/(t_{ex}\!-\bar t)2$ gives (\ref{req})$_2$. 
Eq.  (\ref{req}) is thus explained.

\section{Finite $R$ energies}
\label{Renergies}

\noindent
Using cylindrical coordinates $(y,\rho,\varphi)$ for $\bX'$, one easily finds  
that for  $z_e\!\equiv\!Z\!+\!\Delta\!\le\!0$
the electric force generated by the static charge distribution of fig. \ref{pancake} b),
the associated potential energy $mc^2 \rU$ and the left asymptotes of $\rU$ are
\bea
 \rF(\Delta,\!Z)
\equiv\!\!\int_0^{Z_2(Z)}\!\!\!\!\!\!\!\!dy\, \widetilde{n_0}(y)\!\! \int_0^r\!\!\!d\rho 
\frac {2\pi e^2\rho(y\!-\!z_e)}{\big[\rho{}^2\!+\!\left(y\!-\!z_e\right)^2\big]^{3/2}}\qquad\nn[-6pt]
=\!-\!2\pi e^2\!\!\int_0^{Z_2(Z)}\!\!\!\!\!\!\!\!\!\!\!dy \:\frac { \widetilde{n_0}(y) (y\!-\!z_e)}{
\sqrt{\! (y\!-\!z_e)^2\!+\!r^2}}+4\pi e^2\widetilde{N}\!(\!Z\!). \qquad\qquad  \label{Fzer}\\
\rU(\Delta,\!Z) \equiv\displaystyle\frac {\mu}2  
\displaystyle\int_0^{Z_2(Z)}\!\!\!\!\!\!\!\!\!\!\!\!dy \, \widetilde{n_0}(y) \big[
\sqrt{\! y{}^2\!+\!r^2} \qquad\qquad\qquad\qquad   \label{rU}\\[-6pt]
\qquad\qquad\qquad\qquad \!-\!\sqrt{\! (y\!\!-\!\!Z\!\!-\!\!\Delta)^2\!+\!r^2}\big] 
\!-\!\mu\widetilde{N}\!(\!Z\!)\Delta \!-\! \mu\widetilde{{\cal N}}(Z), \nn
\rU(-\infty,\!Z) \!=\! \displaystyle\frac {\mu}2\!\!
\displaystyle\int_0^{Z_2(Z)}\!\!\!\!\!\!\!\!\!\!\!\!\!\!\!dy \, \widetilde{n_0}\!(\!y\!) \!\!\left[\!\!
\sqrt{\!\! y{}^2\!\!+\!r^2}\!-\!y\right]\!+\!\mu\!\!
\displaystyle\int_0^{Z}\!\!\!\!\!\!\!dy \, \widetilde{n_0}\!(\!y\!) y. 
 \qquad\label{asymptotes}
\eea
Here $\mu\!\equiv\!4\pi e^2\!/mc^2$. $\rU$ is continuous in $(-Z,\!Z)$, since
we have chosen $\U(-Z,Z)$ as the ($\Delta$-independent) 
`additive constants'. Energy conservation implies
$$
\gamma\!+\!\rU(\!\Delta,\!Z)\!=\! \hat\gamma[l,\!Z]\!+\!\rU[\hat \Delta(l,\!Z),\!Z]
\!=\! \hat\gamma[\xi_{ex}(\!Z\!),\!Z]\!+\!\rU(\!-\!Z,\!Z).
$$
The last equality holds only if $\hat z_e(l,\!Z)\!\ge\!0$, i.e. $l\!\le\!\xi_{ex}(Z)$;  
the right-hand side is the electrons' energy  when expelled from the bulk.
This leads to the final relativistic factor
\bea
\rga\!(\! Z\!) \!\!=\! \hat\gamma(\!l\!,\!Z\!)\!-\!\mu\widetilde{N}\!(\!Z\!)\hat z_e\!(\!l\! ,\!Z\!)
\!+\! \displaystyle\frac {\mu}2\!\!\! 
\int_0^{Z_2(Z)}\!\!\!\!\!\!\!\!\!\!\!\!\!\!\!   dy \, \widetilde{n_0}\!(\!y\!) \!\left[\!y\!\!-\!\!
\sqrt{\! [y\!\!-\!\!\hat z_e\!(\!l\!,\!Z\!)]^2\!\!+\!r^2}\!\right]\!\nn[-6pt]
\!=\! \hat\gamma[\xi_{ex}\!(\!Z\!),\!Z]\!+\! \frac {\mu}2\!\!\!
\int_0^{Z_2(Z)}\!\!\!\!\!\!\!\!\!\!\!\!\!\!\!\!dy \, \widetilde{n_0}\!( \!y \!) \! \! \left[y\!-\!
\sqrt{\! y{}^2\!+\!r^2}\right] \! \! \quad\mbox{if }   l\!\le\!\xi_{ex}( \!Z \!).\qquad\:\label{genrga'}
\eea
Deriving this and the identity  $\widetilde{N}[Z_2(Z)]\!=\!2\widetilde{N}(Z)$ we find 
\ $ \widetilde{n_0}[Z_2(Z)]\frac {dZ_2} {dZ}\!=\!2 \widetilde{n_0}(Z)$ \ and that, as claimed, $\rga\!(Z)$ is strictly decreasing,
since $d\rga\!\!/dZ$ is negative-definite:
\bea
\frac {d\rga} {dZ}\!=\!\! \ba{l}
\frac {\partial\hat\gamma(l,\!Z)} {\partial Z}
\!-\!\frac\mu 2\!\frac {dZ_2} {dZ}\widetilde{n_0}[Z_2\!(\!Z\!)]\!\left[\!\sqrt{\! [Z_2\!(\!Z\!)\!-\!\hat z_e\!(\!l\!,\!Z\!)]^2\!+\!r^2}
\!\!-\!\!Z_2\!(\!Z\!)\!\Big]\right.\ea \nn
 \ba{l}\!\!-\! \mu\widetilde{n_0}\!(\!Z\!)\hat z_e\!(\!l\!,\!Z\!)\!-\!\mu\widetilde{N}\!(\!Z\!)
\frac {\partial\hat z_e\!(\!l\!,\!Z\!)} {\partial Z}\!=\!\frac {\partial\hat\gamma} {\partial Z}(\!l\!,\!Z\!)
\!-\!\mu\widetilde{N}\!(\!Z\!)\frac {\partial\hat z_e} {\partial Z}(\!l\!,\!Z\!)\ea\quad\nn
\!\!\!-\!\mu  \widetilde{n_0}(\!Z\!)\!\left[\!\sqrt{\! [Z_2\!(\!Z\!)\!-\!\hat z_e\!(\!l\!,\!Z\!)]^2\!+\!r^2}
\!-\!Z_2\!(\!Z\!)\!+\!\hat z_e\!(\!l\!,\!Z\!)\right]\!\! .
\qquad \quad \label{dgammaZ'}
\eea
For the step-shaped initial density, setting $\phi\!\equiv\!\hat \Delta(l,\!Z)\!-\!Z$,
%
\bea
\ba{l}
\rU(\Delta,\!Z) \!\!=\!\! \frac M4 \!\left[(\Delta\!-\!Z)\sqrt{\! (\Delta\!-\!Z)^2\!+\!r^2}\!-\!4Z(\Delta\!+\!Z) 
\right.\\[6pt]
\!+r^2\sinh^{\scriptscriptstyle -1}\!\!\frac{\Delta\!-\!Z}r\!-\!(\Delta\!+\!Z)\sqrt{\! (\Delta\!+\!Z)^2\!+\!r^2}\!-\!r^2\sinh^{\scriptscriptstyle -1}\!\!\frac{\Delta\!+\!Z}r \\[6pt]
 \left. \quad\qquad \!+2Z^2\!+\!2Z\sqrt{\! 4Z^2\!+\!r^2} \!+\!r^2\sinh^{\scriptscriptstyle -1}\!\frac{2Z}r\right],
\ea\nonumber
\eea
\bea
\rU(-\infty,\!Z)= \ba{l}   \frac {M\!Z}2\!\left[\!\sqrt{\! 4Z^2\!+\!r^2}\!-\!Z
\!+\!\frac {r^2}{2Z}\sinh^{\scriptscriptstyle -1}\!\!\frac{2Z}R\!\right]\!\!,\ea\qquad\qquad \\
\rga\!(Z) \!=\! \hat\gamma(\!l\!,\!Z\!)\!+\!\! \! \ba{l}\frac M4\!\left\{\! \phi\sqrt{\!\phi\!+\!r^2}
\!-\!4Z\hat \Delta(\!l\!,\!Z\!) \!+r^2\sinh^{\scriptscriptstyle -1}\!\frac{\phi}r 
\right. \ea\qquad   \nn
\ba{l} \left.\!-\! [\phi\!+\!2Z]\sqrt{\![\phi\!+\!2Z]^2\!+\!r^2} 
\!- r^2\sinh^{\scriptscriptstyle -1}\!\frac{\phi\!+\!2Z}r \right\}\ea \quad\qquad   \label{rga} \\
\!=\! \hat\gamma[\xi_{ex}\!(\!Z\!)]\!+\!\!\ba{l} M\!\!\left[\!Z^2\!\!-\! \frac  Z2\!\sqrt{\!\! 4Z^2\!\!+\!r^2}
\!-\!\frac{r^2}4\sinh^{\scriptscriptstyle -1}\!\!\frac{2Z}r\!\right] \ea     \:\:\mbox{if }   l\!\le\!\xi_{ex}\!(\!Z\!). \nonumber 
\eea
If $l\!\le\! \xi_{ex}(Z)$ then \
$\partial_{{\scriptscriptstyle Z}}\hat\gamma\!=\!0\!=\!\partial_{{\scriptscriptstyle Z}}\hat\Delta$ at $\xi\!=\! \xi_{ex}(Z)$, \
eq. (\ref{dgammaZ'}) reduces to  \ $ d\rga\!/dZ\!=\!M[Z-\sqrt{\! 4Z^2\!+\!r^2}]$, and (\ref{genspectrum}) to
\bea
 1/\nu(\gamma) =M\,\ZM\big[\sqrt{\!4Z^2\!+\!r^2}\!-\!Z\big]_{Z=\hat Z(\gamma)} .         \label{spectrum}
\eea


\begin{thebibliography}{0}
\bibitem{footnote2}
If $\tau\ll T\!_{{\scriptscriptstyle H}}$, then while the pulse is passing $F_e^z$
can be neglected, and the motion of the electron is close to the one in vacuum; 
the backward acceleration takes place afterwards and is due only to   $F_e^z$, hence the final
energy is smaller. Whereas if \  $\tau\!\gg\! T\!_{{\scriptscriptstyle H}}$ - which was  the standard situation in laboratories 
until a couple of decades ago - then  
$F\!_p v^z$ oscillates many times about 0,  $W\!\simeq\!0$, and the backward acceleration is washed out.
\bibitem{footnote2bis}
Albeit the pump (\ref{pump}) violates the Maxwell equations (due to the $\rho$-dependence),
we adopt  it as for our purposes it is essentially
equivalent to one that fulfills the Maxwell equations and at the time of impact coincides 
with it  in the main part of its support,
while rapidly decaying outside (this and similar schematizations, e.g. the paraxial one, are currently used in the literature).
\bibitem{footnote3}
$\hat u^z,\hat\gamma,\hat \bY,\hat\Xi,\hat\bx_e$ reduce to the 
 $u_e^{z{\scriptscriptstyle (0)}},\gamma_e^{{\scriptscriptstyle (0)}},\bY_e,\Xi_e,\bx_e^{{\scriptscriptstyle (0)}}$  of \cite{Fio14JPA} if  $ \widetilde{n_0}( Z ) \!\equiv\!0$.
\bibitem{footnote6}
There is no need of a recourse to kinetic theory taking
collisions into account, e.g. by BGK \cite{BhaGroKro54} 
equations or effective linear inheritance relations \cite{FioMaiRen14}.
\end{thebibliography}
\end{document}